\documentclass[%
 aip,
 jmp,%
 amsmath,amssymb,
 reprint,%
]{revtex4-1}

\usepackage{graphicx}
\usepackage{dcolumn}
\usepackage{bm}
\usepackage{hyperref}
\usepackage{boxhandler}
\usepackage{float}
\renewcommand{\vec}[1]{\mathbf{#1}}
\let\oldhat\hat
\renewcommand{\hat}[1]{\oldhat{\mathbf{#1}}}
\begin{document}

\title{Role of spin mixing conductance in determining thermal spin pumping near the ferromagnetic phase transition in EuO$_{1-x}$ and La$_{2}$NiMnO$_{6}$ 
}

\author{Kingshuk Mallick}
	\affiliation{Physics Department, Indian Institute of Science, Bangalore - 560012, India}
\author{Aditya Wagh}%
	\affiliation{Physics Department, Indian Institute of Science, Bangalore - 560012, India}%
\author{Adrian Ionescu}%
	\affiliation{Cavendish Laboratory, Physics Department, University of Cambridge, Cambridge CB3 0HE, United Kingdom} 
\author{Crispin H.W. Barnes}%
	\affiliation{Cavendish Laboratory, Physics Department, University of Cambridge, Cambridge CB3 0HE, United Kingdom} 
\author{P.S.Anil Kumar}%
 \email{anil@iisc.ac.in}
\affiliation{Physics Department, Indian Institute of Science, Bangalore - 560012, India}%
\date{\today}
\begin{abstract}
We present a comprehensive study of the temperature (\textit{T}) dependence of the longitudinal spin Seebeck effect (LSSE) in Pt/EuO$_{1-x}$ and Pt/La$_{2}$NiMnO$_{6}$ (LNMO) hybrid structures across their Curie temperatures ($T_c$). Both systems host \textit{ferro}magnetic interaction below $T_c$, hence present optimal conditions for testing magnon spin current based theories against \textit{ferri}magnetic YIG. Notably, we observe an anomalous Nernst effect (ANE) generated voltage in bare EuO$_{1-x}$, however, we find LSSE predominates the thermal signals in the bilayers with Pt. The \textit{T}-dependence of the LSSE in small \textit{T}-range near $T_c$ could be fitted to a power law of the form $(T_{c}-T)^{P}$. The derived critical exponent, \textit{P}, was verified for different methods of LSSE representation and sample crystallinity. The results are explained based on the magnon-driven thermal spin pumping mechanism that relate the \textit{T}-dependence of LSSE to the spin mixing conductance ($G_{mix}$) at the heavy metal/ferromagnet (HM/FM) interface, which in turn is known to vary in accordance with the square of the spontaneous magnetization ($M_s$). Additionally, the \textit{T}-dependence of the real part of $G_{mix}$ derived from spin Hall magnetoresistance measurements at different temperatures for the Pt/LNMO structure, further establish the interdependence.
\end{abstract}

\pacs{Valid PACS appear here}
\keywords{Suggested keywords}
\maketitle

\section{\label{sec:level1}Introduction}
The discovery of spin Seebeck effect (SSE) in 2008 by Uchida \textit{et al.}\cite{uchida2008observation} opened up the multidisciplinary field of spin caloritronics, combining thermoelectricity and spintronics\cite{bauer2012spin, yu2017spin}. In the longitudinal SSE (LSSE) configuration an out-of-plane temperature gradient generates a spin current in a magnetic material which can be detected via the inverse spin Hall effect (ISHE) in an adjoining heavy metal (HM) layer, like Pt and W\cite{uchida2010observation,uchida2014longitudinal}. Another related phenomena observed in ferromagnetic insulator(FI)/HM bilayer is the change in HM resistance depending on the magnetization orientation of the FI layer. A charge current passing through the HM layer can generate a spin current via the spin Hall effect which gets absorbed or reflected from the FI layer depending on the magnetization direction. This modifies the resistance in the HM layer and this phenomena is commonly referred to as the spin Hall magnetoresistance (SMR)\cite{althammer2013quantitative,chen2013theory}. SMR has proven to be a successful approach to quantify the spin mixing interfacial conductance ($G_{mix}$) of FI/NM bilayers \cite{hahn2013comparative,vlietstra2013exchange,velez2016competing,weiler2013experimental}, an important parameter affecting both SMR and LSSE\cite{xiao2010theory,vlietstra2014simultaneous,wang2015spin}.

Temperature variation of LSSE signal has been carried out for investigating various thermospin properties such as phonon-mediated effects\cite{jaworski2010observation,jaworski2011spin,adachi2010gigantic,kikkawa2016magnon,iguchi2017concomitant}, correlation between LSSE and magnon excitation\cite{rezende2014magnon,boona2014magnon,geprags2016origin,jin2015effect}, effects of metal-insulator transition\cite{ramos2013observation,de2019temperature} and recently, antiferromagnetic phase transitions\cite{lin2016enhancement,mallick2019enhanced,prakash2016spin,qiu2018spin,cramer2018spin,baldrati2018spin,khymyn2016transformation,rezende2016diffusive}. The low temperature evolution of LSSE in the prototypical YIG/Pt bilayer far from $T_c$ can be understood based on the magnon spin current theory\cite{rezende2014magnon}. However, when it comes to the temperature dependence near $T_c$, theoretical predictions and experimental evidences have failed to come to a consensus. Uchida et al\cite{uchida2014quantitative} observed a rapid decrease of LSSE signal ($V_{LSSE}$) with an increase in temperature in YIG/Pt, i.e. $V_{LSSE} \propto (T_{c}-T)^{3}$. Measurements on thick films of YIG/Pt by Wang \textit{et al.}\cite{wang2015spin} obtained $V_{LSSE} \propto (T_{c}-T)^{1.5}$. Other than YIG, LSSE(\textit{T}) for two other manganites namely, La$_{0.7}$Sr$_{0.3}$MnO$_{3}$\cite{wu2017longitudinal} and La$_{0.7}$Ca$_{0.3}$MnO$_{3}$\cite{de2019temperature} could be described by, $(T_{c}-T)^{1.9}$. and $(T_{c}-T)^{0.7}$ respectivelty. On the theoretical side, according to the magnon-driven thermal spin pumping mechanism, the LSSE voltage is predominantly determined by $G_{mix}$\cite{xiao2010theory,adachi2013theory,rezende2014magnon} which Ohnuma \textit{et al.}\cite{ohnuma2014enhanced} predicted to follow : $G_{mix} \propto (4 \pi M_{s})^{2}$ near $T_c$. This implies the change in $G_{mix}$ is closely associated with the \textit{T}-dependent magnetic ordering in the sample. Combining these arguments it is expected that $LSSE \propto M_{s}^{2}$, where $M_s$ is the saturation magnetization. Consequently, if \textit{P} and $\beta$ are the critical exponents of LSSE and $M_s$ respectively, then, $P=2\beta$. However, some authors have also presented a different perspective based on numerical and analytical investigations\cite{adachi2018spin,barker2016thermal} which suggest that the LSSE should vary in accordance with the magnetization. Therefore, both should share the same critical exponents. To address these discrepancies from an experimental standpoint, we investigate \textit{T}-evolution of LSSE in EuO$_{1-x}$ and La$_{2}$NiMnO$_{6}$ across their ferromagnet to paramagnet transition temperatures ($T_c$). 

EuO has a rocksalt structure (a = 0.5144 nm)\cite{levy1969spontaneous}, whose large ferromagnetic response, below its curie temperature of 69K, is due to the half-filled 4$f$ Eu$^{2+}$ orbital\cite{dietrich1975spin,kasuya1970exchange,passell1976neutron}. Oxygen deficient EuO, i.e., EuO$_{1-x}$ is intrinsically electron doped which undergoes simultaneous ferromagnetic and insulating-conducting phase transition across which the resistivity can drop by 13 orders of magnitude\cite{oliver1972conductivity,penney1972insulator} and the conduction electrons become nearly 100$\%$ spin polarized\cite{steeneken2002exchange,schmehl2007epitaxial}. Electron doping can also enhance the $T_c$ above 140K\cite{barbagallo2010experimental}. These properties and the close lattice matching with Si makes EuO$_{1-x}$ an excellent candidate for spintronic applications\cite{jansen2012silicon,schmehl2007epitaxial}. EuO has also been predicted to be the ideal candidate to test theories on spin transport across FM/HM bilayers\cite{adachi2018spin}.

La$_{2}$NiMnO$_{6}$ (LNMO) is a double perovskite ferromagnetic insulator which has a Curie temperature close to room temperature ($T_{c} = 280$K)\cite{dass2003oxygen}. Its ferromagnetism arises from $180^{o}$ Ni$^{2+}-O-Mn^{4+}$ superexchange bonding between an empty Mn$^{4+}$ e$_{g}$ orbital and a half-filled d-orbital of the neighboring Ni$^{2+}$ site\cite{das2008electronic,kumar2014temperature}. It is considered to be a promising candidate for  spintronics\cite{guo2006growth,shiomi2014paramagnetic,hashisaka2007spin}. Recently, a spin pumping study from LNMO into Pt by Shiomi and Saitoh\cite{shiomi2014paramagnetic} demonstrated spin transport not only in the ferromagnetic state of LNMO but also in a wide temperature range above $T_c$. This was attributed to short range ferromagnetic correlations that exist in LNMO above $T_c$\cite{zhou2007evidence,guo2009local}. They also present LSSE results in a small temperature range near $T_c$ which is shown to vary in accordance with the magnetization, $M(T)$. In this report, we undertake exhaustive \textit{T}-dependent LSSE measurements on both epitaxial and polcrystalline LNMO films, wherein, we focus on the power law decay of the LSSE signal near $T_c$. Good control of interface quality and optimized measurement conditions ensure a higher signal to noise ratio down to the smallest signal close to $T_c$, thereby allowing direct correlation with $G_{mix}$ obtained from SMR measurements on polycrystalline films. To establish the generality of the observed power law behavior, \textit{T}-dependent LSSE was measured for a polycrystalline Pt/EuO$_{1-x}$ structure as well. 
Interestingly, we observe an anomalous Nernst effect (ANE) signal in EuO$_{1-x}$ without the top Pt. After separating the ANE voltages from the total signal we find that LSSE dominates the electrical signals in Pt/EuO$_{1-x}$. We discuss our results based on the magnon-driven thermal spin pumping mechanism that the relate \textit{T}-evolution of LSSE to $G_{mix}$. 

\begin{figure}[t]
	\centering
	\begin{center}
		\includegraphics[width=0.5\textwidth]{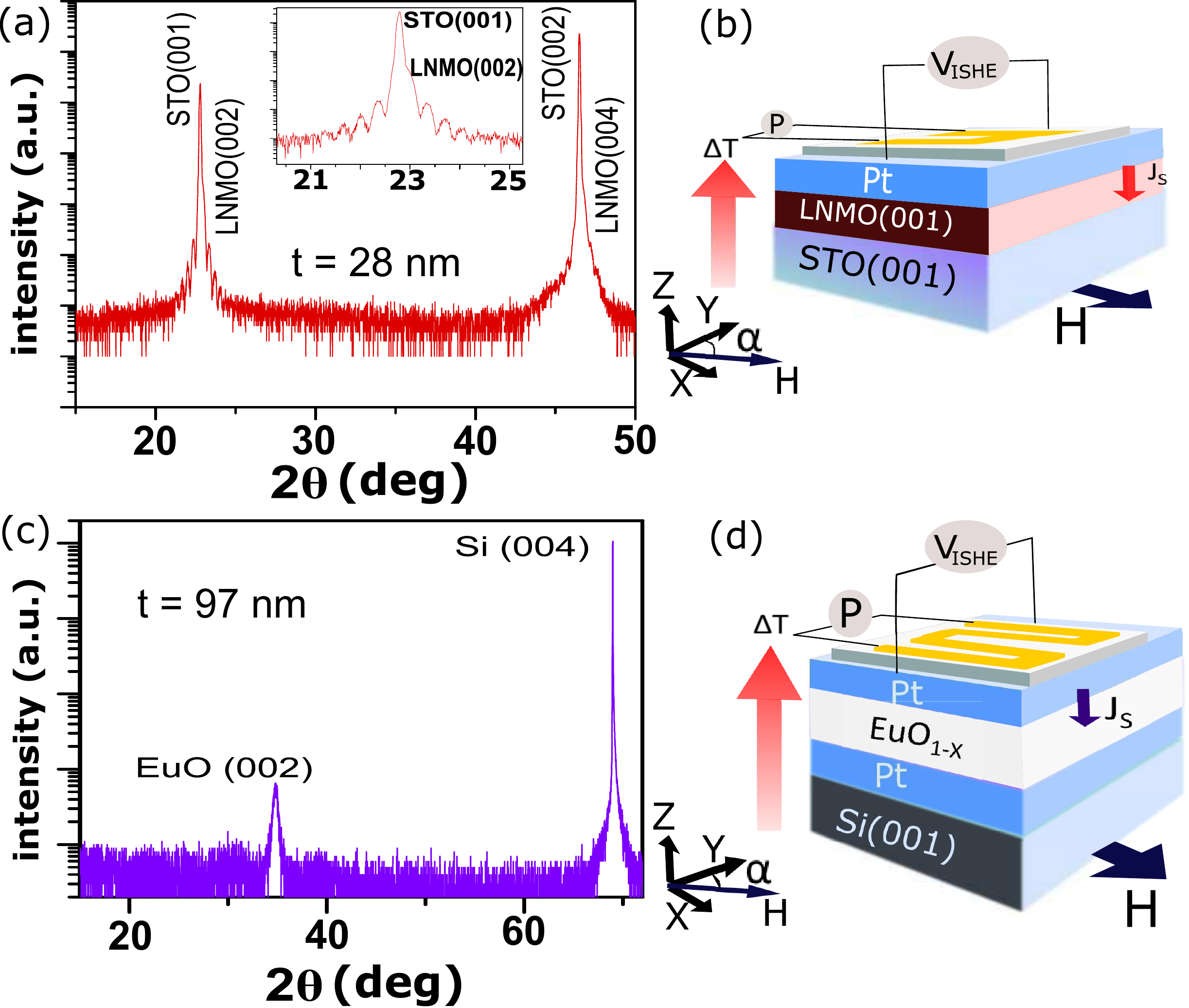}
		\small{\caption{(a) and (c) HR-XRD of the epitaxial LNMO/STO and polycrystalline Pt/EuO/Pt/Si sample respectively. Inset of (a) shows the presence of clear Laue oscillations on either side of the substrate peak. (b) and (d) represent the final device configuration for LSSE measurements\label{fig:xrd}}}
	\end{center}	
\end{figure}

\section{Experiment}
Epitaxial and polycrystalline LNMO films having thicknesses of 57 nm and 28 nm, were grown at 800$^{o}$C and 0.6 mbar O$_{2}$ pressure by pulse laser deposition(PLD) on SrTiO$_{3}$ (001) and Au buffered GGG(111) substrates respectively. Polycrystalline targets were ablated using a KrF laser source with $\lambda = 248$ nm at a repetition rate of 5 Hz. Post deposition, the films were annealed \textit{in-situ} in 500 mbar O$_{2}$ pressure at 600$^{o}$C for 1 hour and subsequently cooled down to room temperature at 5$^{o}$C/min. For LSSE and SMR measurements, Pt was deposited on top of LNMO using a standard e-beam evaporation technique. The surface was cleaned with \textit{in-situ} annealing and Argon plasma before Pt deposition. The nominal sample structure was STO(001)/LNMO(57 nm)/Pt(4.5 nm) (henceforth sample A) and GGG(111)/Au(5 nm)/LNMO(57 nm)/Pt(5 nm) (henceforth sample B).

The polycrystalline EuO$_{1-x}$ sample was deposited at room temperature using a CEVP RF/DC magnetron sputtering system with a base pressure of 5$\times 10^{-9}$ Torr. Co-deposition was performed using two targets: a 99.99\% pure Eu$_2$O$_3$ and a 99.99\% pure Eu target. The EuO$_{1-x}$ film was co-deposited while maintaining the RF power constant at 50 W for the Eu$_2$O$_3$ target and the DC deposition current for the Eu target at 0.15 A. The growth was performed in an Ar$^{+}$ plasma at a pressure of 2 mTorr with a flow rate of 14 sccm. The substrates used were one inch Si (001) with a native oxide layer. One Pt layer was deposited between the substrate and the EuO$_{1-x}$ film and another one on the top at 2 mTorr, with 0.1 A DC current and at 14 sccm Ar flow. The nominal sample structure, Si(001)/SiO$_2$(1.4 nm)/Pt(5 nm)/EuO$_{1-x}$(97 nm)/Pt(5 nm), is shown in Fig. \ref{fig:xrd}(d). The top Pt layer serves as the ISHE detection layer and also protects the EuO$_{1-x}$ from atmospheric degradation. The Pt seed layer was necessary in order to avoid intermixing at the Si/SiO$_2$/EuO$_{1-x}$ interface, which otherwise has resulted in poor EuO$_{1-x}$ films with large roughness.
 
The crystal structure of the films were evaluated by high resolution X-Ray diffraction (HRXRD) using Cu K$\alpha$ radiation. Sample magnetic moment was recorded as a function of field and temperature using SQUID magnetometry. Fig.  \ref{fig:xrd}(a) is the HR-XRD scan on a LNMO(28 nm)/STO sample around the (001)$_{\text{STO}}$ reflections. The pseudocubic pervoskite bulk lattice parameter of LNMO is 3.879${\AA}$\cite{bull2003determination} which is very close to that of STO (= 3.905${\AA}$), hence the LNMO peak appears as humps on the STO peaks. This indicates LNMO films were grown epitaxially which was confirmed from clear Laue fringes around the (002) reflection indicating high crystallinity, flat surface and homogeneity of the grown film (see inset of Fig. \ref{fig:xrd}(a)). In contrast, the EuO$_{1-x}$ on Si (001) has a preferred (001) orientation, as seen in Fig.  \ref{fig:xrd}(c) and is polycrystalline confirmed from the large FWHM of the (002) peak and its omega scan (not shown).

\begin{figure}[t]
	\begin{center}
		\includegraphics[width=0.45\textwidth]{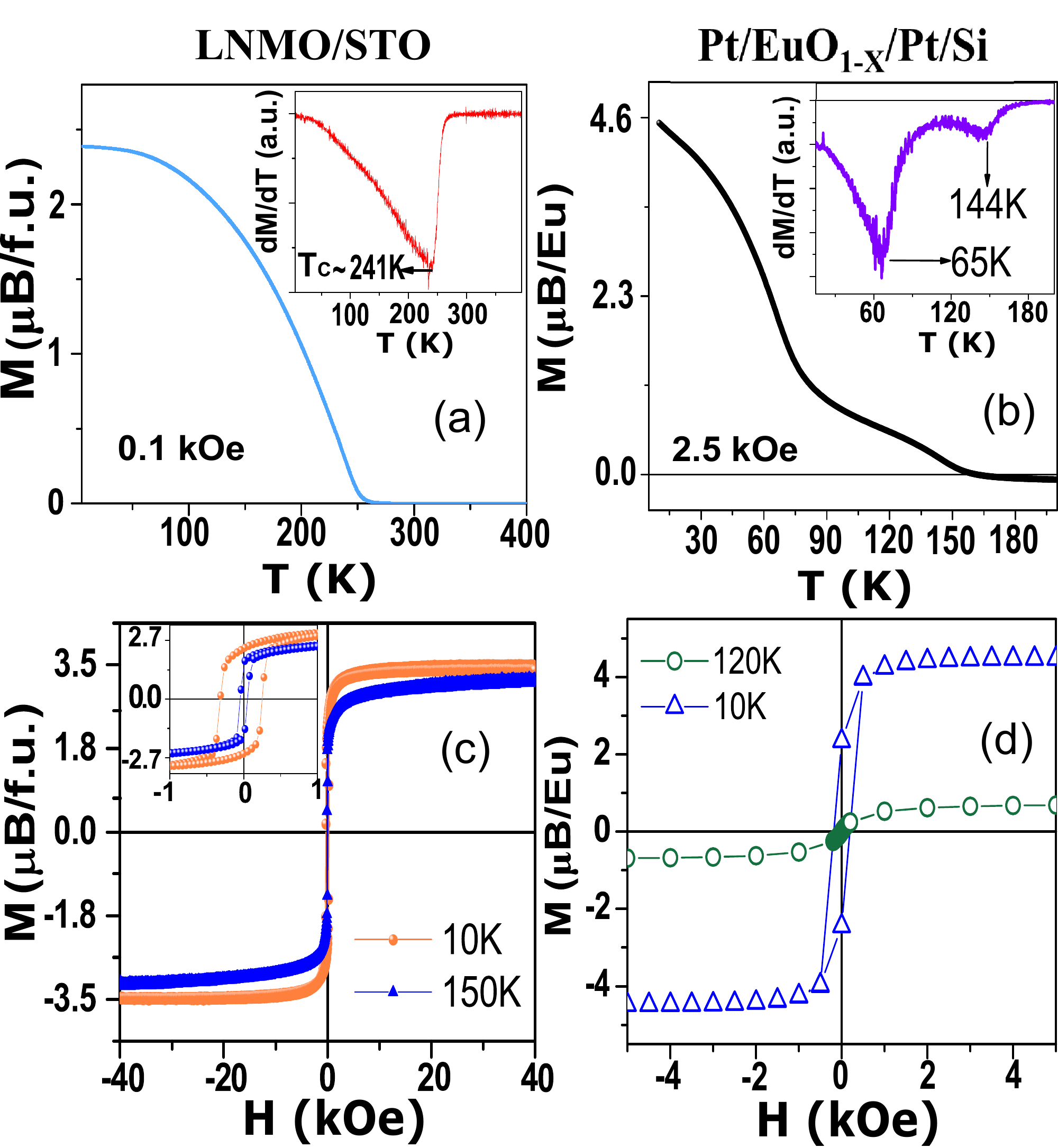}
		\small{\caption{(a) and (b) Field cooled M - T curves of the LNMO/STO and Pt/EuO/Pt/Si sample respectively. Applied in-plane magnetic fields strengths are also denoted. Inset of both figures show presence of minima in the $dM/dT$ curves, depicting the position of $T_c$. (c) and (d) isothermal $M - H$ hysteresis curves with field applied in-plane for LNMO/STO and Pt/EuO/Pt/Si sample respectively. Inset of (c) shows in detail the low field region \label{fig:vsm}}}
	\end{center}	
\end{figure}

The magnetic properties of LNMO thin films including its $T_c$(ferromagnetic to paramagnetic transition) and saturation magnetization ($M_s$) has been found to vary from its bulk ($T_{c}^{bulk}$= 270K, M$_{s}^{bulk}$ at 0K = 5 $\mu_{B}/f.u.$ ) influenced by the growth conditions, film thickness and stoichiometery\cite{bull2003determination,sakurai2011influence,bernal2019non}. The field cooled magnetization (\textit{M}) vs $T$ measured at 100 Oe and its derivative is shown in Fig.  \ref{fig:vsm}(a) and its inset. From the minima in the derivative, we estimate the $T_{c}=$ 241K. $M - H$ loops at $T$ = 10K (Fig. \ref{fig:vsm}(c)) exhibit expected hysteretic behavior with a coercive field and $M_s$ of about 300 Oe and 3.5 $\mu_{B}/f.u.$ respectively. From field cooled $M - T$ of LNMO/Au/GGG (see appendix) the $T_c$ was found identical to LNMO/STO, 241K.

EuO is regarded as a model Heisenberg ferromagnet with $M_s$ at 0K = 7 $\mu_{B}/f.u.$ and $T_c$ = 69K\cite{dietrich1975spin,kasuya1970exchange,passell1976neutron}. The increase in $T_c$ of EuO$_{1-x}$ depends on the extent of electron doping due to O$_{2}$ vacancies\cite{samokhvalov1978nonstoichiometric,barbagallo2010experimental}. The presence of these defects create spin polarized states near the Fermi energy thus modifying the density of states and supplying electrons to the conduction band. The $T_c$ can be enhanced due to conduction-electron-mediated Ruderman-Kittel-Kasuya-Yoshida (RKKY) coupling between the Eu 4$f$ spins\cite{mauger1986magnetic}. Hence a field cooled $M - T$ for the EuO$_{1-x}$ resembles a system having two domes, as observed for our EuO$_{1-x}$ films (Fig.  \ref{fig:vsm}(b))\cite{arnold2008simultaneous,barbagallo2010experimental}. The obtained low temperature feature at 65K ($T_{EuO}$) is close to the bulk value of 69K and another minima at 144K ($T_P$) of the $dM/dT$ curve shown in the inset, is the extended $T_c$ due to RKKY interaction. $M - H$ hysteresis loops of the EuO$_{1-x}$ (Fig.  \ref{fig:vsm}(d)) is characteristic of a soft ferromagnetic film having $M_s$ = 4.6 $\mu_{B}/Eu$ and coercivity less than 80 Oe at 10K. The deviation of $M_s$ from the expected value for a stoichiometric EuO, can be due to the extent of doping, presence of defects or formation of traces of Eu$_{2}$O$_{3}$ upon air exposure.

\begin{figure*}[ht]
		\centering
		\includegraphics[width=0.99\textwidth]{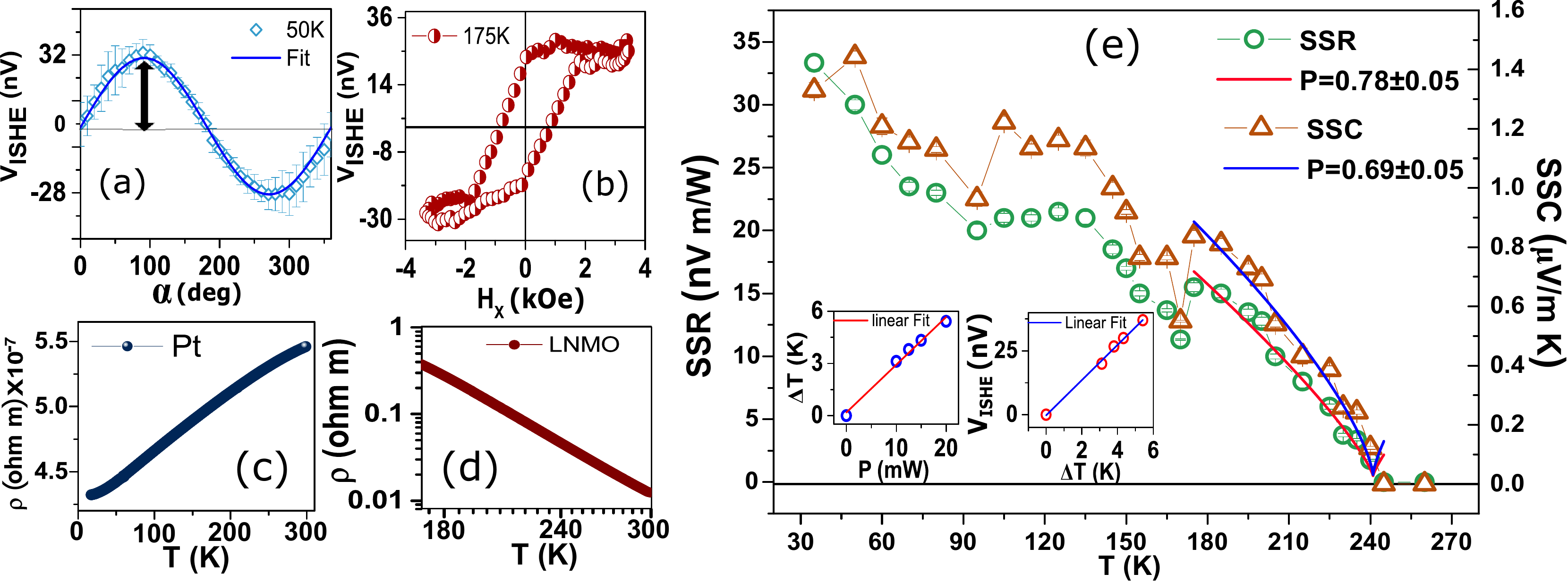}
		\small{\caption{(a) Variation of $V_{ISHE}$ with in-plane angle $\alpha$ at 50K and subsequent fit to a sine function to determine the amplitude, denoted by an arrow. (b) Hysteretic switching of $V_{ISHE}$ as function of in-plane field applied along $\alpha=0$ at 175K. (c) and (d) resistivity as a function of temperature for Pt and LNMO layer respectively, (e) LSSE amplitudes represented as $SSC$ (open triangles) and $SSR$ (open circles) at different temperature and power law fitting near $T_c$. Insets show linear relation between generated voltage, temperature gradient and applied power. \label{fig:lnmo-sto-sse}}}
\end{figure*}

The final sample stack and configuration for LSSE experiments is shown in Fig.  \ref{fig:xrd}(b) and (d). Wire bonding contact was given on the longer edge of the sample to measure the ISHE voltages using a Keithley 2182A nanovoltmeter\cite{uchida2010observation}. LSSE measurements were conducted at different temperature in a modified closed cycle cryostat. To establish a temperature gradient a Cr/Au heater patterned on a sapphire substrate was placed on top of the sample with GE-varnish and a constant small power was applied. This resulted in a perpendicular-to-plane temperature gradient which induced thermal spin currents in the ferromagnetic layer. A constant magnetic field of magnitude 2.5 kOe was rotated in-plane and the change in the generated ISHE voltage was recorded as a function of in-plane angle, $\alpha$,

\begin{equation}
	V_{ISHE} = \rho_{N} \theta_{SHE}  \vec{J_{s}} \times \vec{\sigma}
\end{equation}

where $\theta_{SHE}$ is the spin Hall angle and $\rho_{N}$ is the electrical resistivity of NM layer. The applied field was greater than the anisotropies hence V$_{ISHE}$ has a sinusoidal variation as depicted in Fig.  \ref{fig:lnmo-sto-sse}(a) and Fig.  \ref{fig:EuO-ANE}(a) for 50K and 25K respectively. Similar loops were recorded at different temperature and fitted with a sine function to extract the amplitude (marked with double sided arrow in Fig.  \ref{fig:lnmo-sto-sse}(a)). Field-sweep measurements at $\alpha=90$ have also been carried out at some temperatures which show a hysteretic variation of V$_{ISHE}$ (Fig.  \ref{fig:lnmo-sto-sse}(b) and Fig.  \ref{fig:EuO-ANE}(c)), resembling $M - H$ traces. 
To analyze the temperature dependence of the generated signal it is
important to scale the amplitude either with the temperature gradient, $\Delta T$ (in units of $K$), or with the heat flux, $j_{Q}$ (in units of $W/m^{2}$). $j_{Q}$ can be calculated knowing the applied power and the dimension of the top heater. To acquire $\Delta$\textit{T}, standard Pt thermometery was followed, wherein the Pt resistance is initially calibrated as a function of base temperature which is later utilized to estimate the increase in temperature at the sample surface upon applying a heat flux. Initial studies reported the signal as the spin Seebeck coefficient, $SSC$, where $SSC=V_{ISHE}/(\Delta T\times$L) (in units of $V/Km$), L being the distance between the contacts. However, recently it has become more common to report LSSE as the spin Seebeck resistivity, $SSR$, where $SSR=V_{ISHE}/(j_{Q}\times$L) (in units of $Vm/W$), highlighting the associated errors in the accurate determination of the temperature gradients \cite{sola2017longitudinal,prakash2018evidence}. We report our findings as both $SSC$ and $SSR$ to test the effect of scaling in LSSE analysis. The base temperature was taken from the cryostat's diode sensor reading kept next to the sample. 

\section{LSSE results on LNMO}
First, the two probe resistivity of the LNMO/STO was measured, which was found to be four orders of magnitude larger than that of Pt near $T_c$ (Fig.  \ref{fig:lnmo-sto-sse}(c) and (d)) and demonstrated insulating behavior with temperature. Hence, ANE contributions could be neglected\cite{ramos2013observation}. The heating power was chosen such that it maintains linearity of the V$_{ISHE}$ signal as a function of $\Delta$\textit{T} and power (see inset of Fig.  \ref{fig:lnmo-sto-sse}(e) for 50K data). For a heater of dimension 5 mm$\times$3.2 mm just covering the sample and a distance of 3.7 mm between voltage probes (see Fig.  \ref{fig:xrd}(b)), we show the \textit{T}-evolution of both $SSC$ and $SSR$ for Sample A in Fig.  \ref{fig:lnmo-sto-sse}(e). The signal appears only below $T_c$ (=241K) and then keeps increasing with decrease in temperature till 180K. The $SSR$ at 200K is $1.3 \times 10^{-8}$ in Vm/W which is comparable to the value of $2 \times 10^{-8}$ Vm/W at 200K for Pt(6 nm)/YIG(40 nm) reported by Prakash et al\cite{prakash2018evidence}. In the only other report of LSSE for Pt/LNMO, Shiomi and Saitoh\cite{shiomi2014paramagnetic} show LSSE variation in a small temperature window between 200K and 300K. Although the $\Delta$\textit{T} is mentioned as 10K, the distance between the contacts is not specified. Still, if we assume the the length of the sample (4 mm) as the probing distance, then the $SSC$ at 200K can be approximated as 15 nV/(4 mm$\times$10K) = 0.375 $\mu$V/Km compared to 0.7 $\mu$V/-m obtained in this study. With further decrease in temperature below 180K, the signal initially decreases and then goes through a local maximum around 120K. Below 100K there is again a gradual increase with decrease in \textit{T}. Both $SSC$ and $SSR$ follow the same trend in the entire temperature window except at 30K at which the $SSC$ is seen to drop in contrast to $SSR$. This non-monotonic \textit{T}-dependence can result from a change in thermal magnon parameters like population, conductivity and lifetime\cite{rezende2014magnon,prakash2018evidence} or even effect of interface\cite{guo2016influence} and anisotropy\cite{kalappattil2017roles}. However, in this study, we focus only on the monotonic decrease above 175K that extends up to $T_c$. In analogy to previous reports, this region could be fitted to a ($T_c$-$T$)$^{P}$ power law\cite{wu2017longitudinal,wang2015spin,de2019temperature,uchida2014quantitative}. The derived exponents are $P^{SSR} = 0.78\pm 0.05$ and $P^{SSC} = 0.69\pm 0.05$. 

\begin{figure}[t]
	\begin{center}
		\includegraphics[width=0.4\textwidth]{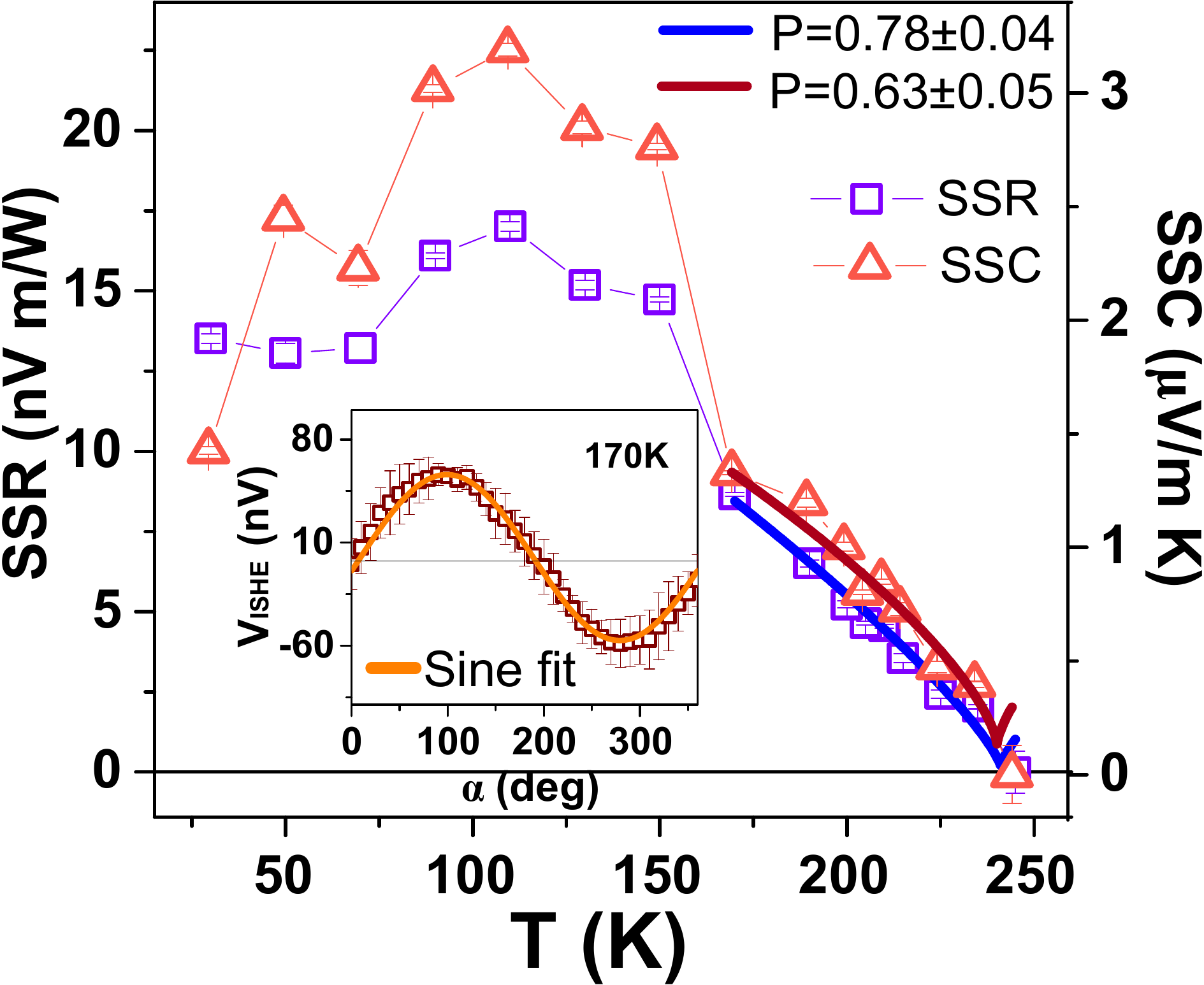}
		\small{\caption{\textit{T}-dependence of LSSE as $SSR$ and $SSC$ representaions in Pt/LNMO/Au/GGG and fit to a power law. Inset shows a typical angular dependence at 170K and corresponding fit to a sine function to extract the amplitude\label{fig:lnmo-ggg-sse}}}
	\end{center}	
\end{figure}

We perform a similar \textit{T}-dependence of LSSE in Sample B to investigate the effect of crystallinity in determining the critical exponent. As expected, the V$_{ISHE}$ displays a sinusoidal variation with in-plane field rotation (see Fig.  \ref{fig:lnmo-ggg-sse} inset) in the entire \textit{T}-range, from which the amplitudes were extracted. We present the \textit{T}-dependence in Fig. \ref{fig:lnmo-ggg-sse} which manifest a similar trend and magnitude as Sample A except at low T. Interestingly, we extract similar critical exponents, $P^{SSR} = 0.78\pm 0.04$ and $P^{SSC} = 0.63\pm 0.05$ by fitting the monotonic decrease in $SSR$ above 170K up to $T_c$(= 241K). This suggests that the same physical mechanism determines spin transport near $T_c$ for both samples irrespective of crystalline order. 

\section{LSSE results on E\lowercase{u}O$_{1-x}$}

\begin{figure}[t]
	\begin{center}
		\includegraphics[width=0.5\textwidth]{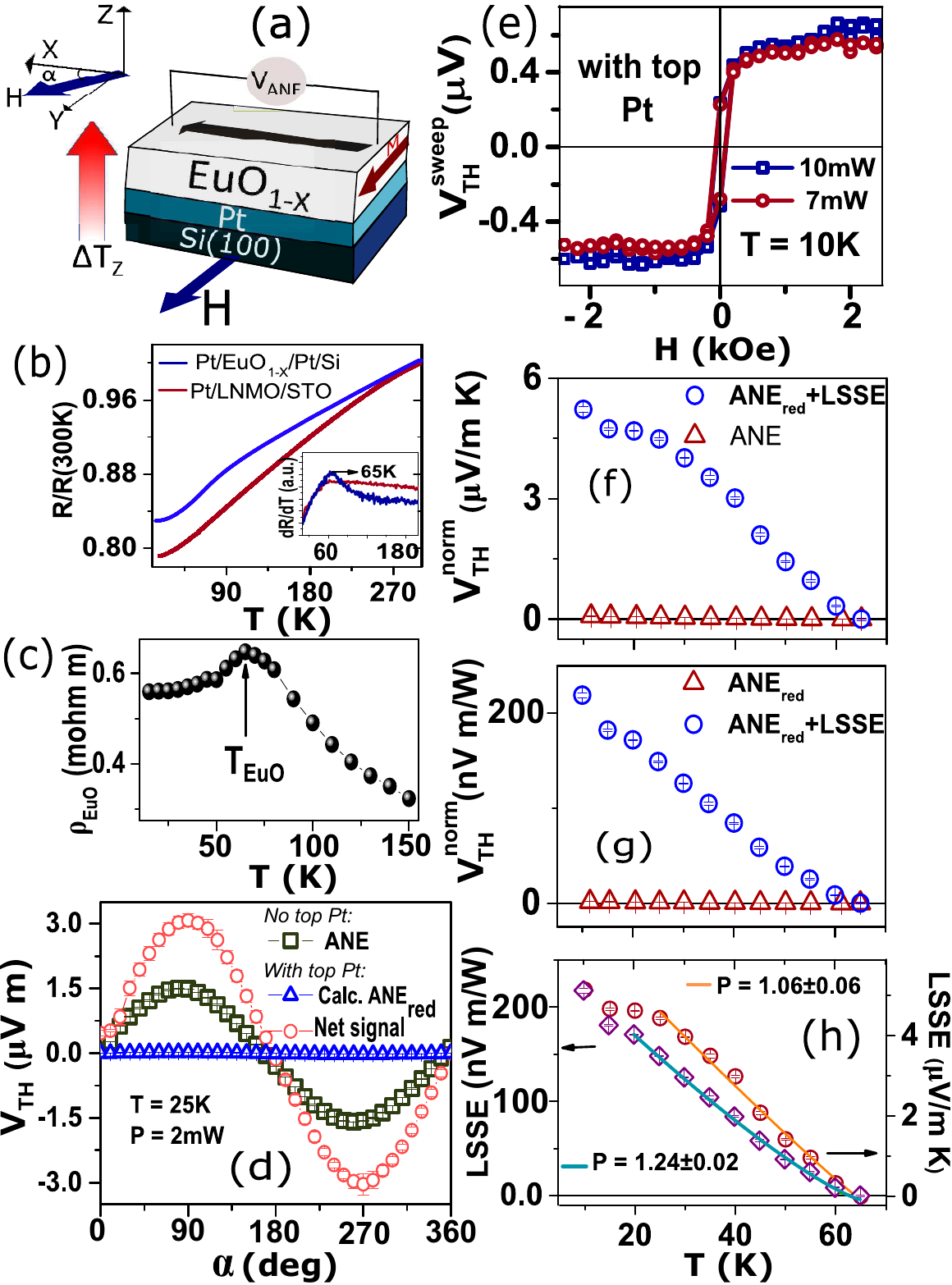}
		\small{\caption{(a) Schematic illustration of the device geometry used for measuring ANE. (b) Variation of stack resistance with temperature for conducting EuO$_{1-x}$ and insulating LNMO. Inset shows the peak in $dR/dT$ for the conducting EuO$_{1-x}$ at $T_c$ of bulk EuO. (c) Calculated \textit{T}-dependence of resistivity of EuO$_{1-x}$ considering a trilayer resistance model. (d) Measured ANE, ANE$_{red}$ and (ANE$_{red}$ + LSSE) voltage in EuO$_{1-x}$ and Pt(5nm)/EuO$_{1-x}$ at 25K as a function of in-plane field angle for a constant power of 2 mW. The values are scaled as (V$_{TH}\times$heater area)/L. (e) Field dependence of the thermal voltage in Pt/EuO$_{1-x}$ for different applied power confirming the ferromagnetic origin of the signal. (f) and (g) \textit{T}-dependence of the reduced ANE voltage for EuO$_{1-x}$ (red triangles) and reduced ANE + LSSE for Pt/EuO$_{1-x}$ (blue circles) in $SSC$ and $SSR$ units respectively. (h) LSSE voltages as $SSC$ and $SSR$ after separation of reduced ANE voltage from the total thermal voltage. Corresponding fits to power law and value of critical exponents are also indicated.\label{fig:EuO-ANE}}}
	\end{center}	
\end{figure}

In principle, the transverse thermal voltage could originate from pure magnon spin currents via LSSE as we observed in LNMO or from spin polarized charge currents via ANE\cite{bougiatioti2017quantitative,ramos2013observation,de2019temperature}. In the case of electron doped EuO$_{1-x}$, there are available defect states in the band gap which allows electron conduction when the	majority states of the spin-split conduction band shift downward to overlap with the defect levels. Hence, generation of a transverse ANE voltage cannot be avoided when a vertical temperature gradient is applied. Accordingly, the ANE voltage is given by:

\begin{equation}
		V_{ANE} = \theta_{ANE}S\hat{m} \times \vec{\Delta T}
\end{equation}
	
where V$_{ANE}$ is the voltage produced by the ANE, $\theta_{ANE}$ is the anomalous Nernst angle, $S$ is the Seebeck coefficient, $\hat{m}$ is the unit vector along magnetization and \textbf{$\Delta$T} is a vector along the temperature gradient. Resistivity determination in bare EuO$_{1-x}$ is tedious owing to the difficulty in getting proper ohmic contacts\cite{altendorf2011oxygen}. However, the conducting nature of our EuO$_{1-x}$ films is evident from the \textit{T}-dependence of Pt/EuO$_{1-x}$/Pt trilayer resistance (see Fig.  \ref{fig:EuO-ANE}(b)) which shows a definite drop at the predicted metal-insulator transition temperature of EuO$_{1-x}$, corresponding to the $T_c$ of bulk EuO (see inset of Fig. \ref{fig:EuO-ANE}(b))\cite{penney1972insulator,oliver1972conductivity}. Hence, a proper analyses of LSSE requires an estimation of the ANE from bare EuO$_{1-x}$. Accordingly, after we measure the \textit{T}-dependence of the LSSE + ANE, \textit{i.e.} the total thermal signal ($V_{TH}$) in the Pt capped sample, we etch away the Pt, followed immediately by a protective coating of GE-varnish. Then we study ANE in the same longitudinal configuration as shown in the schematic of Fig.  \ref{fig:EuO-ANE}(a). The angular variation of ANE with an in-plane applied field displays a similar sinusoidal variation as expected from Equation (2), whose magnitude increases linearly with the applied heater power up to nearly 4mW, as shown in Fig.  \ref{fig:EuO-ANE}(d) and inset. In the same figure, the total thermal signal from a Pt capped EuO$_{1-x}$ is also depicted. It is important to note that the ANE contribution in the total signal would be reduced due to the shunting of currents in the Pt layers, which we represent as ANE$_{red}$. An estimate of the reduction due to the top Pt layer can be done based on the approach by P. Bougiatioti et al\cite{bougiatioti2017quantitative}, who argued that in a NM/FM bilayer, ANE is reduced by a factor $r/(1+r)$, where $r$ is the ratio of electrical conductance, $G$, of FM and NM. Consequently,

\begin{equation}
	r = \frac{G_{EuO_{1-x}}}{G_{Pt}} = \frac{\rho_{Pt}}{\rho_{EuO_{1-x}}}\frac{t_{EuO_{1-x}}}{t_{Pt}}
\end{equation}

with $\rho$ is the resistivity and $t$ the thickness of the corresponding layer. The resistivity of the EuO$_{1-x}$ can be estimated to a fair degree from the measured trilayer resistance, by assuming a parallel connection of three resistances, corresponding to the two Pt layers and the EuO$_{1-x}$ layer (see Fig.\ref{fig:EuO-ANE}(e)). Comparing to other reports on EuO$_{1-x}$\cite{steeneken2012new}, we find that this approach captures the main features of the \textit{T}-dependent resistivity, particularly the MIT at $T_{EuO}$, reasonably well.
 
Substituting these values into Equation (3) along with the measured thicknesses of EuO$_{1-x}$ (97 nm) and Pt (5 nm) we get an estimated 99\% reduction in ANE. Consequently, the thermal signal from the Pt capped sample is predominantly LSSE signal. In addition, the ferromagnetic origin of the thermal signals could be confirmed from the field sweep results,  as depicted in \ref{fig:EuO-ANE}(e), where H is along $\alpha = 0$. Another parasitic voltage that is often associated with Pt, is due to the induced ferromagnetism in Pt, in proximity to a FM. We rule out any significant contribution of this magnetic proximity effect (MPE) in our total thermal signal, based on the results of P. Bougiatioti et al\cite{bougiatioti2017quantitative}, who did not find any MPE in Pt when their FM resistivity was in the same order of magnitude as our EuO$_{1-x}$. Therefore, the pure LSSE signal can be extracted by simply subtracting the reduced ANE (ANE$_{red}$) from the total thermal signal generated from a Pt capped sample. Note that, any LSSE contribution arising at the interface of EuO$_{1-x}$ and bottom Pt layer will be of opposite sign (as the \textit{T}-gradient is reversed) and  very negligible, due to the presence of thick insulating EuO$_{1-x}$ in between.

The \textit{T}-dependence of $ANE_{red}$ and $LSSE+ANE_{red}$ is shown in \ref{fig:EuO-ANE}(f) and (g) for both methods of scaling. An overall decrease in the signal is observed with an increase in \textit{T}, which eventually reduces below the detection limit of our setup ($\sim 10 nV$) above T$_{EuO}$ of 65K.
We extend the same analysis as in LNMO to the pure LSSE signal shown in Fig. \ref{fig:EuO-ANE}(h), by fitting the decay in LSSE signal to a power law of the form ($T_c - T$)$^{P}$. The derived exponents are P$^{SSR} = 1.24\pm 0.02$ and P$^{SSC} = 1.06\pm 0.06$. It is important to note that even though P$^{SSC}$ is less than P$^{SSR}$, resembling the trend in LNMO, the values themselves are higher, arguably closer to one. 

Now we qualitatively discuss the observed power law dependence in LNMO and EuO$_{1-x}$ in accordance with their $M - T$ curve. The magnetization curve of LNMO was analyzed in the critical region by Lou \textit{et al.}\cite{luo2008critical} using the Kouvel-Fisher method that yielded the critical exponent, $P = 0.408 \pm 0.011$. This value was in between those predicted by mean-field model ($=0.5$) and the 3D Heisenberg model ($=0.365$)\cite{stanley1971phase}. A simple power law fitting of our $M$\textit{(T)} data on LNMO/STO also return a similar value of critical exponent, P$_{M(T)^{LNMO}} = 0.39$ (see appendix). Taking the exponent as 0.408, we can now interpret our results based on the magnon-driven thermal spin pumping mechanism. Accordingly, the LSSE(T) should be proportional to (($T_c$-$T$))$^{0.408}$)$^{2}\sim$ ($T_c$-$T$)$^{0.82}$.This is in close agreement with our derived exponents for $P_{SSR}^{LNMO} = 0.78$ in both epitaxial and polycrystalline films and slightly higher than $P_{SSC}^{LNMO}$. 

Stoichiometric EuO is considered an ideal example of a 3D Heisenberg ferromagnet. However, oxygen vacancies in EuO is known to exert a strong influence on its magnetic interactions, thereby increasing the critical exponent to 0.48\cite{gel1975critical}. Such an increment has also been observed for doped EuS\cite{idzuchi2014critical}. The different interactions present in EuO$_{1-x}$ makes the determination of critical exponents non-trivial and hence we adopt the reported value of 0.48 for comparison with our LSSE data. Conforming with our previous arguments, LSSE(\textit{T}) should be proportional to (($T_c$-T))$^{0.48}$)$^{2} = $($T_c - T$)$^{0.96}$ which match quite closely with P$_{SSC}^{EuO_{1-x}} = 1.0$, albeit slightly less than P$_{SSR}^{EuO_{1-x}}$.

Now, the above agreement between the critical exponents of $M(T)$ and LSSE assumes that the dominant \textit{T}-dependent parameter in determining LSSE near $T_c$, is $G_{mix}$. In a simplified picture, one can associate \textit{T}-dependence of $G_{mix}$ solely to its real part, Re[$G_{mix}$], which can be approximated from the \textit{T}-dependence of SMR. Hence, in the next section, we investigate SMR for the polycrystalline LNMO sample.

\section{Spin Hall Magnetoresistance Results}

\begin{figure}[t]
	\begin{center}
		\includegraphics[width=0.5\textwidth]{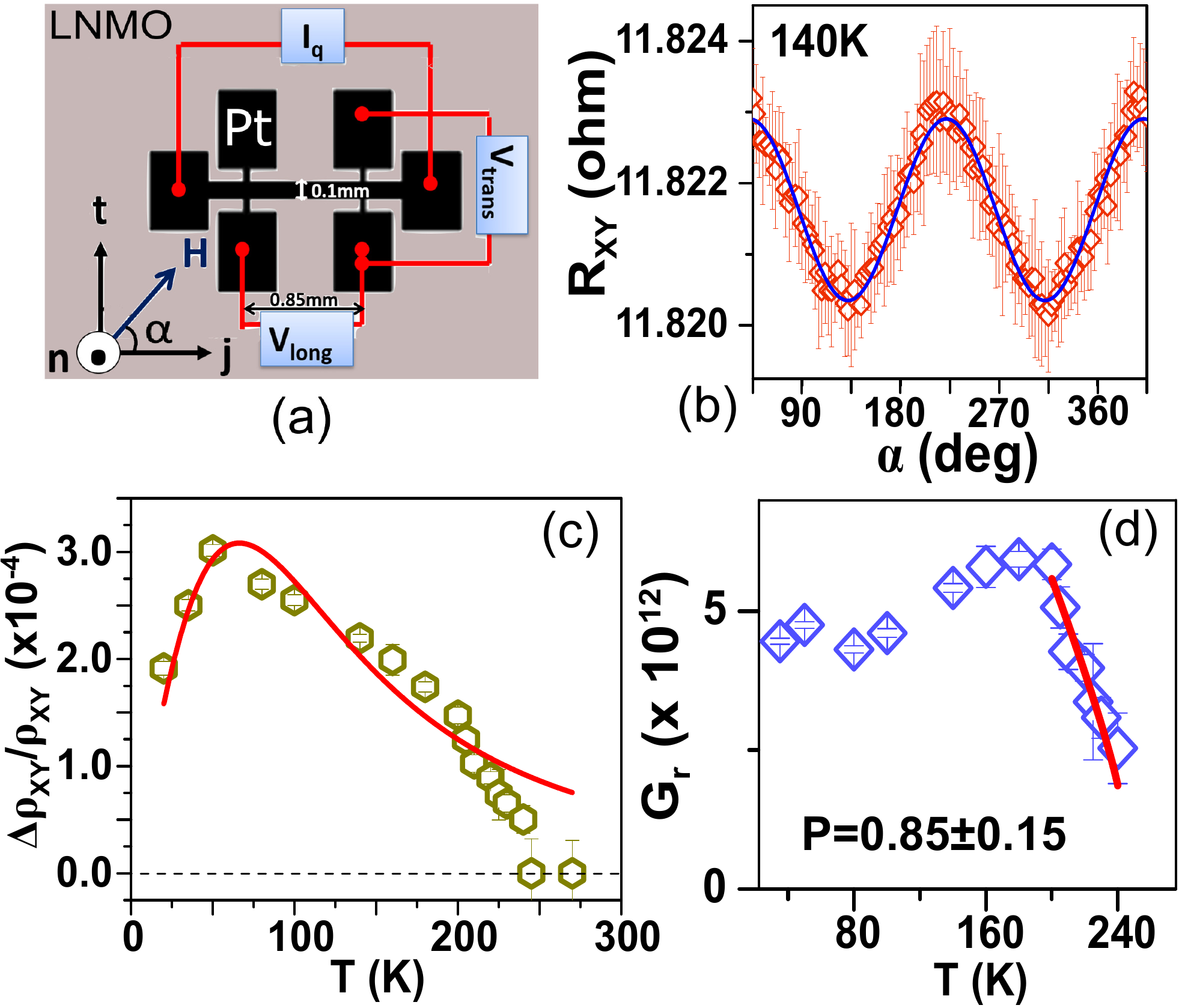}
		\small{\caption{(a) Schematic illustration of the device configuration used for measuring SMR. t, j and n denote the coordinate axes, along, transverse and perpendicular to the current direction, I$_{q}$ respectively. (b) In-plane angular variation of transverse resistance at 140 and 0.25T field. Solid line is fit to $sin(2\alpha)$ to determine SMR amplitude. (c) \textit{T}-dependence of normalized SMR and fit to equation (6) (solid line). (d) Calculated real part of spin mixing conductance at different-\textit{T}. Solid line depicts fit to a power law. Extracted critical exponent is also indicated. \label{fig:LNMO-smr}}}
	\end{center}	
\end{figure}

For SMR measurements we pattern the top Pt layer in Sample B into a Hall bar of dimensions illustrated in \ref{fig:LNMO-smr}(a). A small AC current $\le$100 $\mu$A is applied at 333 Hz frequency and the generated transverse voltage ($V_{trans}$) in Pt is measured using a lock-in amplifier SR830 as a function of in-plane field angle. Here, we utilitize transverse resistivity $\rho_{xy}$ to characterize the SMR due to its low background signal and hence improved signal-to-noise ratio.  \textbf{j} and \textbf{t} denote parallel and transverse to the current direction, whereas \textbf{n} is the out-of-plane direction. $\rho_{xy}$ varies as a function of the magnetization orientation of LNMO, \textbf{m}, as\cite{chen2013theory}:
 \begin{equation}
	\rho_{xy} = \rho_{1}m_{n} + \rho_{2}m_{j}m_{t}
\end{equation}
\begin{equation}
	\rho_{xy} = \rho_{2}\sin(\alpha)\cos(\alpha), \mbox{ for }  m_{n} =0.	
\end{equation}
where $m_{j}, m_{t}$ and $m_{n}$ are projections of \textbf{m} onto the coordinate system, $\alpha$ is the orientation of applied field with respect to the transverse direction and $\rho_{2}$ denote magnitude of reisitivity change due to SMR. Accordingly, Fig. \ref{fig:LNMO-smr}(b) depicts a $\sin(2\alpha)$ variation of $R_{xy}$ and its fit. Parasitic anisotropic magnetoresistance (AMR) contribution arising from conduction in LNMO or magnetic proximity effect (MPE) induced ferromagnetism in Pt is known to satisfy similar symmetry rules as SMR\cite{limmer2006angle}. However, the high resistivity of LNMO compared to Pt prevents any significant current shunting and also avoids MPE induced effects\cite{bougiatioti2017quantitative}. Following the theoretical SMR model described by Chen et al\cite{chen2013theory}, the \textit{T}-dependence of SMR can be depicted using the following ratio,

 \begin{equation}
	\frac{\Delta\rho_{xy}}{\rho_{xy}} = \theta_{sh}^{2}\frac{\lambda_{Pt}(T)}{t_{Pt}}\frac{2\lambda_{Pt}(T)G_{r}tanh^{2}\frac{t_{Pt}}{2\lambda_{Pt}(T)}}{\sigma_{Pt}(T)+2\lambda_{Pt}(T)G_{r}coth\frac{t_{Pt}}{\lambda_{Pt}(T)}}. 
\end{equation}

where $\theta_{sh}$, $t_{Pt}$, $\lambda_{Pt}(T)$ and $\sigma_{Pt}(T)$ are the spin Hall angle, thickness, spin diffusion length and conductivity of Pt, respectively and $G_{r}$ is real part of G$_{mix}$. We illustrate the \textit{T}-dependence of SMR ratio in Fig. \ref{fig:LNMO-smr}(c). Interestingly, the signal exhibits a peak around 50K and vanishes above T$_{c} = 241K$. This suggests that just like LSSE, SMR is also regulated by the long range ferromagnetic correlations in the sample and is not affected by the short range interactions that is known to exist in LNMO even above $T_c$\cite{uchida2015spin}.
 
The \textit{T}-dependent parameters in the above equation have been extensively investigated by different groups. For instance Marmion \textit{et al.}\cite{marmion2014temperature} ascribed the \textit{T}-dependence of SMR to variation in $\lambda_{Pt}(T)$, determined by the Elliot-Yafet mechanism for spin relaxation. $\theta_{sh}$ on the other hand is reported to change very weakly above 100K hence is often taken as constant\cite{isasa2015temperature}. We try fitting our SMR data based on Equation (6), assuming a \textit{T}-independent $G_{r}$ and a $\theta_{sh}=$0.08, only varying $\lambda_{Pt}(T)$ according to the Elliot-Yafet mechanism ($\lambda_{Pt}(T) = C/T$ in unit of nm). However, as discussed by Wang \textit{et al.}\cite{wang2015spin} for Pt/YIG, it fails to reproduce the high-\textit{T} data (see fit in Fig. \ref{fig:LNMO-smr}(c)). Hence, we consider a \textit{T}-dependent $G_{r}(T)$ which can be quantified by rearranging Equation (6) as follows\cite{wang2015spin}:

\begin{equation}
	G_{r}(T) = \frac{\sigma_{Pt}(T)}{2\lambda_{Pt}(T)\Bigg[\frac{\theta_{sh}^2\frac{\lambda_{Pt}(T)}{t_{Pt}}tanh^2\frac{t_{Pt}}{2\lambda_{Pt}(T)}}{\frac{\Delta\rho_{xy}}{\rho_{xy}}}-coth\frac{t_{Pt}}{\lambda_{Pt}(T)}\Bigg]} 
\end{equation}

We consider $G_r(T)$ to be a function of \textit{T} and back-calculate its values from our experimentally measured SMR results using Equation 7, adopting \text{$\theta_{sh}$} = 0.08 and $\lambda_{Pt}(T) = (2.6 \times 10^{-7})/T$ in unit of nm\cite{marmion2014temperature,isasa2015temperature,wang2015spin} (refined from previous fit). We plot the calculated $G_{r}(T)$ and the power law fit in Fig. \ref{fig:LNMO-smr}(d). The important trait we observe is that, the critical exponents conform very nearly to that of LSSE and also with the spin pumping mechanism outlined by Ohnuma \textit{et al.}\cite{ohnuma2014enhanced} (see Table \ref{Table 1}) . It was speculated that previous investigations of critical exponents on YIG failed to come to a consensus with the theoretical predictions because of the \textit{ferri}magnetic nature of YIG, or other considerations like the magnetic surface anisotropy\cite{adachi2018spin}. In this report, both samples exhibit \textit{ferro}magnetic interactions which allow direct comparison with the magnon-driven spin current model\cite{xiao2010theory}. We also acknowledge that our choice of FM allows analysis adopting this simple picture, wherein we only consider \textit{T}-dependence of interfacial spin conductance and disregard other parameters such as the bulk spin conductance\cite{okamoto2016spin} and magnon chemical potential\cite{cornelissen2016magnon}, which can also affect LSSE. Additionally, we could verify the interdependence of $G_{r}(T)$ and LSSE(T) without including the effective spin conductance, $G_s$\cite{velez2018spin}, in Equation (6) suggesting negligible contribution from this term for LNMO. Incorporating these effects might better reproduce the behavior near $T_c$ for other systems.

In LSSE, the signals are generated by magnon spin currents at the bulk of the material which travel to the interface and get pumped into the Pt layer. Hence both bulk and interface magnetization can affect the generated signals. In our experiments we observed that the power law exponents of the SSE are related to the power law exponent of the volume magnetization recorded using a standard SQUID magnetometer. Nature of \textit{M} vs \textit{T} and this correlation suggests that the interface magnetization contribution if present is identical to the bulk. Lastly, we can comment on the impact of scaling, namely $SSR$ or $SSC$, on the derived exponents. We found a better correspondence between P$^{SSR}$ and P$^{M_{s}^2}$ for LNMO and in contrast P$^{SSC}$ conformed better with P$^{M_{s}^2}$ for EuO$_{1-x}$. This might suggest that at low T, when heat transport properties such as thermal conductivity and specific heat undergo large changes, $SSC$ would be a better representation to incorporate those changes. However, near room temperature, $SSR$ representation overcomes uncertainties due to parasitic temperature drops across various interfaces, hence might serve as a better choice. Alternately, one can also argue that, Pt spin conversion parameters, especially $\lambda_{Pt}(T)$, which is known to increase appreciably only below 100K, also affects the LSSE signal and hence needs to be accounted for in the analysis. Simultaneous measurements of all \textit{T}-dependent parameters at low-\textit{T} could be helpful in resolving this question. 

\begin{table}
\centering
\begin{tabular}{ c c c c c c } 
 \hline
  & P$^{SSR}$ & P$^{SSC}$ & P$^{SMR}$ & $T_c$ (K) & P$_{M-T}$ \\
 \hline	      
 Sample A & 0.78$\pm$0.05 & 0.69$\pm$0.05 & - & 241 & 0.41 \\
 Sample B & 0.78$\pm$0.04 & 0.63$\pm$0.05 & 0.85$\pm$0.15 & 241 & 0.41 \\ 
 EuO$_{1-x}$ & 1.24$\pm$0.02 & 1.06$\pm$0.06 & - & 65 & 0.48 \\
 \hline
\end{tabular}
\caption{List of samples and corresponding refined and adopted parameters.}
\label{Table 1}
\end{table}

\section{Conclusion}
The \textit{T}-dependence of LSSE has been studied for three different Pt/FM hybrid structures across its ferromagnet to paramagnet transition temperature, namely, Pt/LNMO/STO, Pt/LNMO/Au/GGG and Pt/EuO$_{1-x}$/Pt/Si. Pure LSSE signal was obtained from the highly resistive LNMO whereas the LSSE had to be disentangled from the ANE signal generated in conducting EuO$_{1-x}$. A power law behavior could describe the decay in LSSE approaching $T_c$ for both LNMO and EuO$_{1-x}$, but the derived critical exponents were found to be characteristic of the material. We could interpret this power law behavior based on the magnon-driven thermal spin pumping mechanism which suggest G$_{mix}$ is the dominating parameter affecting LSSE and which in turn is proportional to $M_{s}^{2}$. Additionally, we show this evaluation remains invariant despite varying the crystalline order in LNMO, but the method used for scaling LSSE becomes important, especially at low-\textit{T}. Finally, we confirm the correlation between magnetization and $G_{mix}$ from SMR measurements on Pt/LNMO at different temperatures. Our work establishes the importance of $G_{mix}$ in determining LSSE across ferromagnetic phase transition and also highlights the correlation between critical exponents of magnetic order parameter and thermal spin transport across NM/FM interfaces. Further systematic studies on different samples having different thicknesses and interface conditions are necessary to confirm whether the exponent is material specific or not. However, this correlation strongly suggests that for materials like LNMO and EuO$_{1-x}$, having simple magnetic structures, $G_{mix}$ is the dominant parameter affecting LSSE. This serves as an important benchmark for future investigations.

\section{Acknowledgement}
We would like to thank Dr. Timo Kuschel for fruitful discussions on anomalous Nernst contributions, Sukanya Pal for crystallographic structure analysis of LNMO, Suresh Pittala for providing the PLD ablation target and Manasa for SQUID measurements. K.M. would like to thank MHRD for financial support. P.S. Anil Kumar would like to thank DST Nano Mission for financial support. 
\vfill	
\section{appendix}
\begin{figure}[H]
	\begin{center}
		\includegraphics[width=0.4\textwidth]{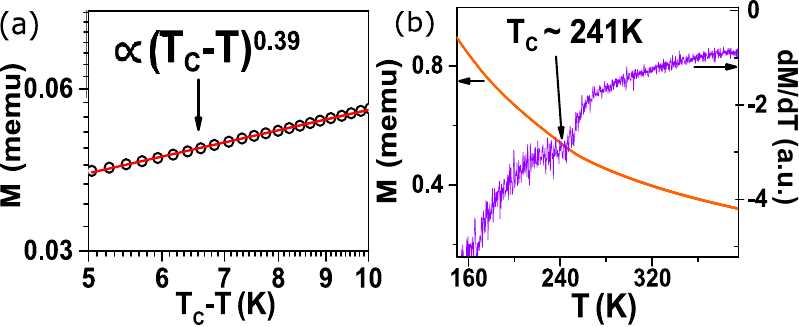}
		\small{\caption{(a) A double logarithmic plot of $T_c - T$ dependence of the magnetization for epitaxial LNMO. (b) $M$ vs $T$ at 100 Oe and its derivative for polycrystalline LNMO sample depicting the position of $T_c$. The GGG substrate contributes to the large paramagnetic background.\label{fig:LNMO-fit}}}
	\end{center}	
\end{figure}

In Fig. \ref{fig:LNMO-fit}(a) we depict a simplified approach towards estimating the critical exponent of magnetization by fitting the \textit{M-T} of epitaxial LNMO near \text{$T_c$}. It would have been ideal to compare the value of P obtained from \textit{M} vs \textit{T} and $SSR$, $SSC$ vs \textit{T} in the same temperature range very close to \text{$T_c$}, but the nature of the spin Seebeck experiments prevent this direct comparison. Ensuring that we operate in the linear region (as shown in Fig. \ref{fig:lnmo-sto-sse}(e)) and generate a minimum measurable signal of ~tens of nV (limited by the experimental setup), we are required to maintain a $\Delta T$ between 4K - 6K near \text{$T_c$}. In addition, since the change in signal with temperature is not large, a minimum step size of 5K was chosen to properly resolve the signals. These limitations meant that in the same temperature range as \textit{M} vs \textit{T}, we had only two spin Seebeck data points. Therefore, to incorporate more data points a larger range was taken. In Fig. \ref{fig:LNMO-fit}(b), we highlight the position of the Curie temperature for the polycrystalline LNMO sample from the derivative of its \textit{M-T} curve.


\begin{thebibliography}{81}%
\makeatletter
\providecommand \@ifxundefined [1]{%
 \@ifx{#1\undefined}
}%
\providecommand \@ifnum [1]{%
 \ifnum #1\expandafter \@firstoftwo
 \else \expandafter \@secondoftwo
 \fi
}%
\providecommand \@ifx [1]{%
 \ifx #1\expandafter \@firstoftwo
 \else \expandafter \@secondoftwo
 \fi
}%
\providecommand \natexlab [1]{#1}%
\providecommand \enquote  [1]{``#1''}%
\providecommand \bibnamefont  [1]{#1}%
\providecommand \bibfnamefont [1]{#1}%
\providecommand \citenamefont [1]{#1}%
\providecommand \href@noop [0]{\@secondoftwo}%
\providecommand \href [0]{\begingroup \@sanitize@url \@href}%
\providecommand \@href[1]{\@@startlink{#1}\@@href}%
\providecommand \@@href[1]{\endgroup#1\@@endlink}%
\providecommand \@sanitize@url [0]{\catcode `\\12\catcode `\$12\catcode
  `\&12\catcode `\#12\catcode `\^12\catcode `\_12\catcode `\%12\relax}%
\providecommand \@@startlink[1]{}%
\providecommand \@@endlink[0]{}%
\providecommand \url  [0]{\begingroup\@sanitize@url \@url }%
\providecommand \@url [1]{\endgroup\@href {#1}{\urlprefix }}%
\providecommand \urlprefix  [0]{URL }%
\providecommand \Eprint [0]{\href }%
\providecommand \doibase [0]{http://dx.doi.org/}%
\providecommand \selectlanguage [0]{\@gobble}%
\providecommand \bibinfo  [0]{\@secondoftwo}%
\providecommand \bibfield  [0]{\@secondoftwo}%
\providecommand \translation [1]{[#1]}%
\providecommand \BibitemOpen [0]{}%
\providecommand \bibitemStop [0]{}%
\providecommand \bibitemNoStop [0]{.\EOS\space}%
\providecommand \EOS [0]{\spacefactor3000\relax}%
\providecommand \BibitemShut  [1]{\csname bibitem#1\endcsname}%
\let\auto@bib@innerbib\@empty
\bibitem [{\citenamefont {Uchida}\ \emph {et~al.}(2008)\citenamefont {Uchida},
  \citenamefont {Takahashi}, \citenamefont {Harii}, \citenamefont {Ieda},
  \citenamefont {Koshibae}, \citenamefont {Ando}, \citenamefont {Maekawa},\
  and\ \citenamefont {Saitoh}}]{uchida2008observation}%
  \BibitemOpen
  \bibfield  {author} {\bibinfo {author} {\bibfnamefont {K.}~\bibnamefont
  {Uchida}}, \bibinfo {author} {\bibfnamefont {S.}~\bibnamefont {Takahashi}},
  \bibinfo {author} {\bibfnamefont {K.}~\bibnamefont {Harii}}, \bibinfo
  {author} {\bibfnamefont {J.}~\bibnamefont {Ieda}}, \bibinfo {author}
  {\bibfnamefont {W.}~\bibnamefont {Koshibae}}, \bibinfo {author}
  {\bibfnamefont {K.}~\bibnamefont {Ando}}, \bibinfo {author} {\bibfnamefont
  {S.}~\bibnamefont {Maekawa}}, \ and\ \bibinfo {author} {\bibfnamefont
  {E.}~\bibnamefont {Saitoh}},\ }\bibfield  {title} {\enquote {\bibinfo {title}
  {Observation of the spin seebeck effect},}\ }\href@noop {} {\bibfield
  {journal} {\bibinfo  {journal} {Nature}\ }\textbf {\bibinfo {volume} {455}},\
  \bibinfo {pages} {778} (\bibinfo {year} {2008})}\BibitemShut {NoStop}%
\bibitem [{\citenamefont {Bauer}, \citenamefont {Saitoh},\ and\ \citenamefont
  {Van~Wees}(2012)}]{bauer2012spin}%
  \BibitemOpen
  \bibfield  {author} {\bibinfo {author} {\bibfnamefont {G.~E.}\ \bibnamefont
  {Bauer}}, \bibinfo {author} {\bibfnamefont {E.}~\bibnamefont {Saitoh}}, \
  and\ \bibinfo {author} {\bibfnamefont {B.~J.}\ \bibnamefont {Van~Wees}},\
  }\bibfield  {title} {\enquote {\bibinfo {title} {Spin caloritronics},}\
  }\href@noop {} {\bibfield  {journal} {\bibinfo  {journal} {Nature materials}\
  }\textbf {\bibinfo {volume} {11}},\ \bibinfo {pages} {391} (\bibinfo {year}
  {2012})}\BibitemShut {NoStop}%
\bibitem [{\citenamefont {Yu}, \citenamefont {Brechet},\ and\ \citenamefont
  {Ansermet}(2017)}]{yu2017spin}%
  \BibitemOpen
  \bibfield  {author} {\bibinfo {author} {\bibfnamefont {H.}~\bibnamefont
  {Yu}}, \bibinfo {author} {\bibfnamefont {S.~D.}\ \bibnamefont {Brechet}}, \
  and\ \bibinfo {author} {\bibfnamefont {J.-P.}\ \bibnamefont {Ansermet}},\
  }\bibfield  {title} {\enquote {\bibinfo {title} {Spin caloritronics, origin
  and outlook},}\ }\href@noop {} {\bibfield  {journal} {\bibinfo  {journal}
  {Physics Letters A}\ }\textbf {\bibinfo {volume} {381}},\ \bibinfo {pages}
  {825--837} (\bibinfo {year} {2017})}\BibitemShut {NoStop}%
\bibitem [{\citenamefont {Uchida}\ \emph {et~al.}(2010)\citenamefont {Uchida},
  \citenamefont {Adachi}, \citenamefont {Ota}, \citenamefont {Nakayama},
  \citenamefont {Maekawa},\ and\ \citenamefont
  {Saitoh}}]{uchida2010observation}%
  \BibitemOpen
  \bibfield  {author} {\bibinfo {author} {\bibfnamefont {K.-i.}\ \bibnamefont
  {Uchida}}, \bibinfo {author} {\bibfnamefont {H.}~\bibnamefont {Adachi}},
  \bibinfo {author} {\bibfnamefont {T.}~\bibnamefont {Ota}}, \bibinfo {author}
  {\bibfnamefont {H.}~\bibnamefont {Nakayama}}, \bibinfo {author}
  {\bibfnamefont {S.}~\bibnamefont {Maekawa}}, \ and\ \bibinfo {author}
  {\bibfnamefont {E.}~\bibnamefont {Saitoh}},\ }\bibfield  {title} {\enquote
  {\bibinfo {title} {Observation of longitudinal spin-seebeck effect in
  magnetic insulators},}\ }\href@noop {} {\bibfield  {journal} {\bibinfo
  {journal} {Applied Physics Letters}\ }\textbf {\bibinfo {volume} {97}},\
  \bibinfo {pages} {172505} (\bibinfo {year} {2010})}\BibitemShut {NoStop}%
\bibitem [{\citenamefont {Uchida}\ \emph
  {et~al.}(2014{\natexlab{a}})\citenamefont {Uchida}, \citenamefont {Ishida},
  \citenamefont {Kikkawa}, \citenamefont {Kirihara}, \citenamefont {Murakami},\
  and\ \citenamefont {Saitoh}}]{uchida2014longitudinal}%
  \BibitemOpen
  \bibfield  {author} {\bibinfo {author} {\bibfnamefont {K.}~\bibnamefont
  {Uchida}}, \bibinfo {author} {\bibfnamefont {M.}~\bibnamefont {Ishida}},
  \bibinfo {author} {\bibfnamefont {T.}~\bibnamefont {Kikkawa}}, \bibinfo
  {author} {\bibfnamefont {A.}~\bibnamefont {Kirihara}}, \bibinfo {author}
  {\bibfnamefont {T.}~\bibnamefont {Murakami}}, \ and\ \bibinfo {author}
  {\bibfnamefont {E.}~\bibnamefont {Saitoh}},\ }\bibfield  {title} {\enquote
  {\bibinfo {title} {Longitudinal spin seebeck effect: from fundamentals to
  applications},}\ }\href@noop {} {\bibfield  {journal} {\bibinfo  {journal}
  {Journal of Physics: Condensed Matter}\ }\textbf {\bibinfo {volume} {26}},\
  \bibinfo {pages} {343202} (\bibinfo {year} {2014}{\natexlab{a}})}\BibitemShut
  {NoStop}%
\bibitem [{\citenamefont {Althammer}\ \emph {et~al.}(2013)\citenamefont
  {Althammer}, \citenamefont {Meyer}, \citenamefont {Nakayama}, \citenamefont
  {Schreier}, \citenamefont {Altmannshofer}, \citenamefont {Weiler},
  \citenamefont {Huebl}, \citenamefont {Gepr{\"a}gs}, \citenamefont {Opel},
  \citenamefont {Gross} \emph {et~al.}}]{althammer2013quantitative}%
  \BibitemOpen
  \bibfield  {author} {\bibinfo {author} {\bibfnamefont {M.}~\bibnamefont
  {Althammer}}, \bibinfo {author} {\bibfnamefont {S.}~\bibnamefont {Meyer}},
  \bibinfo {author} {\bibfnamefont {H.}~\bibnamefont {Nakayama}}, \bibinfo
  {author} {\bibfnamefont {M.}~\bibnamefont {Schreier}}, \bibinfo {author}
  {\bibfnamefont {S.}~\bibnamefont {Altmannshofer}}, \bibinfo {author}
  {\bibfnamefont {M.}~\bibnamefont {Weiler}}, \bibinfo {author} {\bibfnamefont
  {H.}~\bibnamefont {Huebl}}, \bibinfo {author} {\bibfnamefont
  {S.}~\bibnamefont {Gepr{\"a}gs}}, \bibinfo {author} {\bibfnamefont
  {M.}~\bibnamefont {Opel}}, \bibinfo {author} {\bibfnamefont {R.}~\bibnamefont
  {Gross}},  \emph {et~al.},\ }\bibfield  {title} {\enquote {\bibinfo {title}
  {Quantitative study of the spin hall magnetoresistance in ferromagnetic
  insulator/normal metal hybrids},}\ }\href@noop {} {\bibfield  {journal}
  {\bibinfo  {journal} {Physical Review B}\ }\textbf {\bibinfo {volume} {87}},\
  \bibinfo {pages} {224401} (\bibinfo {year} {2013})}\BibitemShut {NoStop}%
\bibitem [{\citenamefont {Chen}\ \emph {et~al.}(2013)\citenamefont {Chen},
  \citenamefont {Takahashi}, \citenamefont {Nakayama}, \citenamefont
  {Althammer}, \citenamefont {Goennenwein}, \citenamefont {Saitoh},\ and\
  \citenamefont {Bauer}}]{chen2013theory}%
  \BibitemOpen
  \bibfield  {author} {\bibinfo {author} {\bibfnamefont {Y.-T.}\ \bibnamefont
  {Chen}}, \bibinfo {author} {\bibfnamefont {S.}~\bibnamefont {Takahashi}},
  \bibinfo {author} {\bibfnamefont {H.}~\bibnamefont {Nakayama}}, \bibinfo
  {author} {\bibfnamefont {M.}~\bibnamefont {Althammer}}, \bibinfo {author}
  {\bibfnamefont {S.~T.}\ \bibnamefont {Goennenwein}}, \bibinfo {author}
  {\bibfnamefont {E.}~\bibnamefont {Saitoh}}, \ and\ \bibinfo {author}
  {\bibfnamefont {G.~E.}\ \bibnamefont {Bauer}},\ }\bibfield  {title} {\enquote
  {\bibinfo {title} {Theory of spin {H}all magnetoresistance},}\ }\href@noop {}
  {\bibfield  {journal} {\bibinfo  {journal} {Physical Review B}\ }\textbf
  {\bibinfo {volume} {87}},\ \bibinfo {pages} {144411} (\bibinfo {year}
  {2013})}\BibitemShut {NoStop}%
\bibitem [{\citenamefont {Hahn}\ \emph {et~al.}(2013)\citenamefont {Hahn},
  \citenamefont {De~Loubens}, \citenamefont {Klein}, \citenamefont {Viret},
  \citenamefont {Naletov},\ and\ \citenamefont
  {Youssef}}]{hahn2013comparative}%
  \BibitemOpen
  \bibfield  {author} {\bibinfo {author} {\bibfnamefont {C.}~\bibnamefont
  {Hahn}}, \bibinfo {author} {\bibfnamefont {G.}~\bibnamefont {De~Loubens}},
  \bibinfo {author} {\bibfnamefont {O.}~\bibnamefont {Klein}}, \bibinfo
  {author} {\bibfnamefont {M.}~\bibnamefont {Viret}}, \bibinfo {author}
  {\bibfnamefont {V.~V.}\ \bibnamefont {Naletov}}, \ and\ \bibinfo {author}
  {\bibfnamefont {J.~B.}\ \bibnamefont {Youssef}},\ }\bibfield  {title}
  {\enquote {\bibinfo {title} {Comparative measurements of inverse spin {H}all
  effects and magnetoresistance in {YIG}/{P}t and {YIG}/{T}a},}\ }\href@noop {}
  {\bibfield  {journal} {\bibinfo  {journal} {Physical Review B}\ }\textbf
  {\bibinfo {volume} {87}},\ \bibinfo {pages} {174417} (\bibinfo {year}
  {2013})}\BibitemShut {NoStop}%
\bibitem [{\citenamefont {Vlietstra}\ \emph {et~al.}(2013)\citenamefont
  {Vlietstra}, \citenamefont {Shan}, \citenamefont {Castel}, \citenamefont
  {Ben~Youssef}, \citenamefont {Bauer},\ and\ \citenamefont
  {Van~Wees}}]{vlietstra2013exchange}%
  \BibitemOpen
  \bibfield  {author} {\bibinfo {author} {\bibfnamefont {N.}~\bibnamefont
  {Vlietstra}}, \bibinfo {author} {\bibfnamefont {J.}~\bibnamefont {Shan}},
  \bibinfo {author} {\bibfnamefont {V.}~\bibnamefont {Castel}}, \bibinfo
  {author} {\bibfnamefont {J.}~\bibnamefont {Ben~Youssef}}, \bibinfo {author}
  {\bibfnamefont {G.}~\bibnamefont {Bauer}}, \ and\ \bibinfo {author}
  {\bibfnamefont {B.}~\bibnamefont {Van~Wees}},\ }\bibfield  {title} {\enquote
  {\bibinfo {title} {Exchange magnetic field torques in {YIG}/{P}t bilayers
  observed by the spin-{H}all magnetoresistance},}\ }\href@noop {} {\bibfield
  {journal} {\bibinfo  {journal} {Applied Physics Letters}\ }\textbf {\bibinfo
  {volume} {103}},\ \bibinfo {pages} {032401} (\bibinfo {year}
  {2013})}\BibitemShut {NoStop}%
\bibitem [{\citenamefont {V{\'e}lez}\ \emph {et~al.}(2016)\citenamefont
  {V{\'e}lez}, \citenamefont {Bedoya-Pinto}, \citenamefont {Yan}, \citenamefont
  {Hueso},\ and\ \citenamefont {Casanova}}]{velez2016competing}%
  \BibitemOpen
  \bibfield  {author} {\bibinfo {author} {\bibfnamefont {S.}~\bibnamefont
  {V{\'e}lez}}, \bibinfo {author} {\bibfnamefont {A.}~\bibnamefont
  {Bedoya-Pinto}}, \bibinfo {author} {\bibfnamefont {W.}~\bibnamefont {Yan}},
  \bibinfo {author} {\bibfnamefont {L.~E.}\ \bibnamefont {Hueso}}, \ and\
  \bibinfo {author} {\bibfnamefont {F.}~\bibnamefont {Casanova}},\ }\bibfield
  {title} {\enquote {\bibinfo {title} {Competing effects at {P}t/{YIG}
  interfaces: Spin {H}all magnetoresistance, magnon excitations, and magnetic
  frustration},}\ }\href@noop {} {\bibfield  {journal} {\bibinfo  {journal}
  {Physical Review B}\ }\textbf {\bibinfo {volume} {94}},\ \bibinfo {pages}
  {174405} (\bibinfo {year} {2016})}\BibitemShut {NoStop}%
\bibitem [{\citenamefont {Weiler}\ \emph {et~al.}(2013)\citenamefont {Weiler},
  \citenamefont {Althammer}, \citenamefont {Schreier}, \citenamefont {Lotze},
  \citenamefont {Pernpeintner}, \citenamefont {Meyer}, \citenamefont {Huebl},
  \citenamefont {Gross}, \citenamefont {Kamra}, \citenamefont {Xiao} \emph
  {et~al.}}]{weiler2013experimental}%
  \BibitemOpen
  \bibfield  {author} {\bibinfo {author} {\bibfnamefont {M.}~\bibnamefont
  {Weiler}}, \bibinfo {author} {\bibfnamefont {M.}~\bibnamefont {Althammer}},
  \bibinfo {author} {\bibfnamefont {M.}~\bibnamefont {Schreier}}, \bibinfo
  {author} {\bibfnamefont {J.}~\bibnamefont {Lotze}}, \bibinfo {author}
  {\bibfnamefont {M.}~\bibnamefont {Pernpeintner}}, \bibinfo {author}
  {\bibfnamefont {S.}~\bibnamefont {Meyer}}, \bibinfo {author} {\bibfnamefont
  {H.}~\bibnamefont {Huebl}}, \bibinfo {author} {\bibfnamefont
  {R.}~\bibnamefont {Gross}}, \bibinfo {author} {\bibfnamefont
  {A.}~\bibnamefont {Kamra}}, \bibinfo {author} {\bibfnamefont
  {J.}~\bibnamefont {Xiao}},  \emph {et~al.},\ }\bibfield  {title} {\enquote
  {\bibinfo {title} {Experimental test of the spin mixing interface
  conductivity concept},}\ }\href@noop {} {\bibfield  {journal} {\bibinfo
  {journal} {Physical review letters}\ }\textbf {\bibinfo {volume} {111}},\
  \bibinfo {pages} {176601} (\bibinfo {year} {2013})}\BibitemShut {NoStop}%
\bibitem [{\citenamefont {Xiao}\ \emph {et~al.}(2010)\citenamefont {Xiao},
  \citenamefont {Bauer}, \citenamefont {Uchida}, \citenamefont {Saitoh},
  \citenamefont {Maekawa} \emph {et~al.}}]{xiao2010theory}%
  \BibitemOpen
  \bibfield  {author} {\bibinfo {author} {\bibfnamefont {J.}~\bibnamefont
  {Xiao}}, \bibinfo {author} {\bibfnamefont {G.~E.}\ \bibnamefont {Bauer}},
  \bibinfo {author} {\bibfnamefont {K.-c.}\ \bibnamefont {Uchida}}, \bibinfo
  {author} {\bibfnamefont {E.}~\bibnamefont {Saitoh}}, \bibinfo {author}
  {\bibfnamefont {S.}~\bibnamefont {Maekawa}},  \emph {et~al.},\ }\bibfield
  {title} {\enquote {\bibinfo {title} {Theory of magnon-driven spin seebeck
  effect},}\ }\href@noop {} {\bibfield  {journal} {\bibinfo  {journal}
  {Physical Review B}\ }\textbf {\bibinfo {volume} {81}},\ \bibinfo {pages}
  {214418} (\bibinfo {year} {2010})}\BibitemShut {NoStop}%
\bibitem [{\citenamefont {Vlietstra}\ \emph {et~al.}(2014)\citenamefont
  {Vlietstra}, \citenamefont {Shan}, \citenamefont {Van~Wees}, \citenamefont
  {Isasa}, \citenamefont {Casanova},\ and\ \citenamefont
  {Youssef}}]{vlietstra2014simultaneous}%
  \BibitemOpen
  \bibfield  {author} {\bibinfo {author} {\bibfnamefont {N.}~\bibnamefont
  {Vlietstra}}, \bibinfo {author} {\bibfnamefont {J.}~\bibnamefont {Shan}},
  \bibinfo {author} {\bibfnamefont {B.}~\bibnamefont {Van~Wees}}, \bibinfo
  {author} {\bibfnamefont {M.}~\bibnamefont {Isasa}}, \bibinfo {author}
  {\bibfnamefont {F.}~\bibnamefont {Casanova}}, \ and\ \bibinfo {author}
  {\bibfnamefont {J.~B.}\ \bibnamefont {Youssef}},\ }\bibfield  {title}
  {\enquote {\bibinfo {title} {Simultaneous detection of the spin-hall
  magnetoresistance and the spin-seebeck effect in platinum and tantalum on
  yttrium iron garnet},}\ }\href@noop {} {\bibfield  {journal} {\bibinfo
  {journal} {Physical Review B}\ }\textbf {\bibinfo {volume} {90}},\ \bibinfo
  {pages} {174436} (\bibinfo {year} {2014})}\BibitemShut {NoStop}%
\bibitem [{\citenamefont {Wang}\ \emph {et~al.}(2015)\citenamefont {Wang},
  \citenamefont {Zou}, \citenamefont {Zhang}, \citenamefont {Cai},
  \citenamefont {Wang}, \citenamefont {Shen},\ and\ \citenamefont
  {Sun}}]{wang2015spin}%
  \BibitemOpen
  \bibfield  {author} {\bibinfo {author} {\bibfnamefont {S.}~\bibnamefont
  {Wang}}, \bibinfo {author} {\bibfnamefont {L.}~\bibnamefont {Zou}}, \bibinfo
  {author} {\bibfnamefont {X.}~\bibnamefont {Zhang}}, \bibinfo {author}
  {\bibfnamefont {J.}~\bibnamefont {Cai}}, \bibinfo {author} {\bibfnamefont
  {S.}~\bibnamefont {Wang}}, \bibinfo {author} {\bibfnamefont {B.}~\bibnamefont
  {Shen}}, \ and\ \bibinfo {author} {\bibfnamefont {J.}~\bibnamefont {Sun}},\
  }\bibfield  {title} {\enquote {\bibinfo {title} {Spin {S}eebeck effect and
  spin {H}all magnetoresistance at high temperatures for a {P}t/yttrium iron
  garnet hybrid structure},}\ }\href@noop {} {\bibfield  {journal} {\bibinfo
  {journal} {Nanoscale}\ }\textbf {\bibinfo {volume} {7}},\ \bibinfo {pages}
  {17812--17819} (\bibinfo {year} {2015})}\BibitemShut {NoStop}%
\bibitem [{\citenamefont {Jaworski}\ \emph {et~al.}(2010)\citenamefont
  {Jaworski}, \citenamefont {Yang}, \citenamefont {Mack}, \citenamefont
  {Awschalom}, \citenamefont {Heremans},\ and\ \citenamefont
  {Myers}}]{jaworski2010observation}%
  \BibitemOpen
  \bibfield  {author} {\bibinfo {author} {\bibfnamefont {C.}~\bibnamefont
  {Jaworski}}, \bibinfo {author} {\bibfnamefont {J.}~\bibnamefont {Yang}},
  \bibinfo {author} {\bibfnamefont {S.}~\bibnamefont {Mack}}, \bibinfo {author}
  {\bibfnamefont {D.}~\bibnamefont {Awschalom}}, \bibinfo {author}
  {\bibfnamefont {J.}~\bibnamefont {Heremans}}, \ and\ \bibinfo {author}
  {\bibfnamefont {R.}~\bibnamefont {Myers}},\ }\bibfield  {title} {\enquote
  {\bibinfo {title} {Observation of the spin-seebeck effect in a ferromagnetic
  semiconductor},}\ }\href@noop {} {\bibfield  {journal} {\bibinfo  {journal}
  {Nature materials}\ }\textbf {\bibinfo {volume} {9}},\ \bibinfo {pages} {898}
  (\bibinfo {year} {2010})}\BibitemShut {NoStop}%
\bibitem [{\citenamefont {Jaworski}\ \emph {et~al.}(2011)\citenamefont
  {Jaworski}, \citenamefont {Yang}, \citenamefont {Mack}, \citenamefont
  {Awschalom}, \citenamefont {Myers},\ and\ \citenamefont
  {Heremans}}]{jaworski2011spin}%
  \BibitemOpen
  \bibfield  {author} {\bibinfo {author} {\bibfnamefont {C.}~\bibnamefont
  {Jaworski}}, \bibinfo {author} {\bibfnamefont {J.}~\bibnamefont {Yang}},
  \bibinfo {author} {\bibfnamefont {S.}~\bibnamefont {Mack}}, \bibinfo {author}
  {\bibfnamefont {D.}~\bibnamefont {Awschalom}}, \bibinfo {author}
  {\bibfnamefont {R.}~\bibnamefont {Myers}}, \ and\ \bibinfo {author}
  {\bibfnamefont {J.}~\bibnamefont {Heremans}},\ }\bibfield  {title} {\enquote
  {\bibinfo {title} {Spin-seebeck effect: A phonon driven spin distribution},}\
  }\href@noop {} {\bibfield  {journal} {\bibinfo  {journal} {Physical review
  letters}\ }\textbf {\bibinfo {volume} {106}},\ \bibinfo {pages} {186601}
  (\bibinfo {year} {2011})}\BibitemShut {NoStop}%
\bibitem [{\citenamefont {Adachi}\ \emph {et~al.}(2010)\citenamefont {Adachi},
  \citenamefont {Uchida}, \citenamefont {Saitoh}, \citenamefont {Ohe},
  \citenamefont {Takahashi},\ and\ \citenamefont
  {Maekawa}}]{adachi2010gigantic}%
  \BibitemOpen
  \bibfield  {author} {\bibinfo {author} {\bibfnamefont {H.}~\bibnamefont
  {Adachi}}, \bibinfo {author} {\bibfnamefont {K.-i.}\ \bibnamefont {Uchida}},
  \bibinfo {author} {\bibfnamefont {E.}~\bibnamefont {Saitoh}}, \bibinfo
  {author} {\bibfnamefont {J.-i.}\ \bibnamefont {Ohe}}, \bibinfo {author}
  {\bibfnamefont {S.}~\bibnamefont {Takahashi}}, \ and\ \bibinfo {author}
  {\bibfnamefont {S.}~\bibnamefont {Maekawa}},\ }\bibfield  {title} {\enquote
  {\bibinfo {title} {Gigantic enhancement of spin seebeck effect by phonon
  drag},}\ }\href@noop {} {\bibfield  {journal} {\bibinfo  {journal} {Applied
  Physics Letters}\ }\textbf {\bibinfo {volume} {97}},\ \bibinfo {pages}
  {252506} (\bibinfo {year} {2010})}\BibitemShut {NoStop}%
\bibitem [{\citenamefont {Kikkawa}\ \emph {et~al.}(2016)\citenamefont
  {Kikkawa}, \citenamefont {Shen}, \citenamefont {Flebus}, \citenamefont
  {Duine}, \citenamefont {Uchida}, \citenamefont {Qiu}, \citenamefont {Bauer},\
  and\ \citenamefont {Saitoh}}]{kikkawa2016magnon}%
  \BibitemOpen
  \bibfield  {author} {\bibinfo {author} {\bibfnamefont {T.}~\bibnamefont
  {Kikkawa}}, \bibinfo {author} {\bibfnamefont {K.}~\bibnamefont {Shen}},
  \bibinfo {author} {\bibfnamefont {B.}~\bibnamefont {Flebus}}, \bibinfo
  {author} {\bibfnamefont {R.~A.}\ \bibnamefont {Duine}}, \bibinfo {author}
  {\bibfnamefont {K.-i.}\ \bibnamefont {Uchida}}, \bibinfo {author}
  {\bibfnamefont {Z.}~\bibnamefont {Qiu}}, \bibinfo {author} {\bibfnamefont
  {G.~E.}\ \bibnamefont {Bauer}}, \ and\ \bibinfo {author} {\bibfnamefont
  {E.}~\bibnamefont {Saitoh}},\ }\bibfield  {title} {\enquote {\bibinfo {title}
  {Magnon polarons in the spin seebeck effect},}\ }\href@noop {} {\bibfield
  {journal} {\bibinfo  {journal} {Physical review letters}\ }\textbf {\bibinfo
  {volume} {117}},\ \bibinfo {pages} {207203} (\bibinfo {year}
  {2016})}\BibitemShut {NoStop}%
\bibitem [{\citenamefont {Iguchi}\ \emph {et~al.}(2017)\citenamefont {Iguchi},
  \citenamefont {Uchida}, \citenamefont {Daimon},\ and\ \citenamefont
  {Saitoh}}]{iguchi2017concomitant}%
  \BibitemOpen
  \bibfield  {author} {\bibinfo {author} {\bibfnamefont {R.}~\bibnamefont
  {Iguchi}}, \bibinfo {author} {\bibfnamefont {K.-i.}\ \bibnamefont {Uchida}},
  \bibinfo {author} {\bibfnamefont {S.}~\bibnamefont {Daimon}}, \ and\ \bibinfo
  {author} {\bibfnamefont {E.}~\bibnamefont {Saitoh}},\ }\bibfield  {title}
  {\enquote {\bibinfo {title} {Concomitant enhancement of the longitudinal spin
  seebeck effect and the thermal conductivity in a {P}t/{YIG}/{P}t system at
  low temperatures},}\ }\href@noop {} {\bibfield  {journal} {\bibinfo
  {journal} {Physical Review B}\ }\textbf {\bibinfo {volume} {95}},\ \bibinfo
  {pages} {174401} (\bibinfo {year} {2017})}\BibitemShut {NoStop}%
\bibitem [{\citenamefont {Rezende}\ \emph {et~al.}(2014)\citenamefont
  {Rezende}, \citenamefont {Rodr{\'\i}guez-Su{\'a}rez}, \citenamefont {Cunha},
  \citenamefont {Rodrigues}, \citenamefont {Machado}, \citenamefont {Guerra},
  \citenamefont {Ortiz},\ and\ \citenamefont {Azevedo}}]{rezende2014magnon}%
  \BibitemOpen
  \bibfield  {author} {\bibinfo {author} {\bibfnamefont {S.}~\bibnamefont
  {Rezende}}, \bibinfo {author} {\bibfnamefont {R.}~\bibnamefont
  {Rodr{\'\i}guez-Su{\'a}rez}}, \bibinfo {author} {\bibfnamefont
  {R.}~\bibnamefont {Cunha}}, \bibinfo {author} {\bibfnamefont
  {A.}~\bibnamefont {Rodrigues}}, \bibinfo {author} {\bibfnamefont
  {F.}~\bibnamefont {Machado}}, \bibinfo {author} {\bibfnamefont {G.~F.}\
  \bibnamefont {Guerra}}, \bibinfo {author} {\bibfnamefont {J.~L.}\
  \bibnamefont {Ortiz}}, \ and\ \bibinfo {author} {\bibfnamefont
  {A.}~\bibnamefont {Azevedo}},\ }\bibfield  {title} {\enquote {\bibinfo
  {title} {Magnon spin-current theory for the longitudinal spin-seebeck
  effect},}\ }\href@noop {} {\bibfield  {journal} {\bibinfo  {journal}
  {Physical Review B}\ }\textbf {\bibinfo {volume} {89}},\ \bibinfo {pages}
  {014416} (\bibinfo {year} {2014})}\BibitemShut {NoStop}%
\bibitem [{\citenamefont {Boona}\ and\ \citenamefont
  {Heremans}(2014)}]{boona2014magnon}%
  \BibitemOpen
  \bibfield  {author} {\bibinfo {author} {\bibfnamefont {S.~R.}\ \bibnamefont
  {Boona}}\ and\ \bibinfo {author} {\bibfnamefont {J.~P.}\ \bibnamefont
  {Heremans}},\ }\bibfield  {title} {\enquote {\bibinfo {title} {Magnon thermal
  mean free path in yttrium iron garnet},}\ }\href@noop {} {\bibfield
  {journal} {\bibinfo  {journal} {Physical Review B}\ }\textbf {\bibinfo
  {volume} {90}},\ \bibinfo {pages} {064421} (\bibinfo {year}
  {2014})}\BibitemShut {NoStop}%
\bibitem [{\citenamefont {Gepr{\"a}gs}\ \emph {et~al.}(2016)\citenamefont
  {Gepr{\"a}gs}, \citenamefont {Kehlberger}, \citenamefont {Della~Coletta},
  \citenamefont {Qiu}, \citenamefont {Guo}, \citenamefont {Schulz},
  \citenamefont {Mix}, \citenamefont {Meyer}, \citenamefont {Kamra},
  \citenamefont {Althammer} \emph {et~al.}}]{geprags2016origin}%
  \BibitemOpen
  \bibfield  {author} {\bibinfo {author} {\bibfnamefont {S.}~\bibnamefont
  {Gepr{\"a}gs}}, \bibinfo {author} {\bibfnamefont {A.}~\bibnamefont
  {Kehlberger}}, \bibinfo {author} {\bibfnamefont {F.}~\bibnamefont
  {Della~Coletta}}, \bibinfo {author} {\bibfnamefont {Z.}~\bibnamefont {Qiu}},
  \bibinfo {author} {\bibfnamefont {E.-J.}\ \bibnamefont {Guo}}, \bibinfo
  {author} {\bibfnamefont {T.}~\bibnamefont {Schulz}}, \bibinfo {author}
  {\bibfnamefont {C.}~\bibnamefont {Mix}}, \bibinfo {author} {\bibfnamefont
  {S.}~\bibnamefont {Meyer}}, \bibinfo {author} {\bibfnamefont
  {A.}~\bibnamefont {Kamra}}, \bibinfo {author} {\bibfnamefont
  {M.}~\bibnamefont {Althammer}},  \emph {et~al.},\ }\bibfield  {title}
  {\enquote {\bibinfo {title} {Origin of the spin seebeck effect in compensated
  ferrimagnets},}\ }\href@noop {} {\bibfield  {journal} {\bibinfo  {journal}
  {Nature communications}\ }\textbf {\bibinfo {volume} {7}},\ \bibinfo {pages}
  {10452} (\bibinfo {year} {2016})}\BibitemShut {NoStop}%
\bibitem [{\citenamefont {Jin}\ \emph {et~al.}(2015)\citenamefont {Jin},
  \citenamefont {Boona}, \citenamefont {Yang}, \citenamefont {Myers},\ and\
  \citenamefont {Heremans}}]{jin2015effect}%
  \BibitemOpen
  \bibfield  {author} {\bibinfo {author} {\bibfnamefont {H.}~\bibnamefont
  {Jin}}, \bibinfo {author} {\bibfnamefont {S.~R.}\ \bibnamefont {Boona}},
  \bibinfo {author} {\bibfnamefont {Z.}~\bibnamefont {Yang}}, \bibinfo {author}
  {\bibfnamefont {R.~C.}\ \bibnamefont {Myers}}, \ and\ \bibinfo {author}
  {\bibfnamefont {J.~P.}\ \bibnamefont {Heremans}},\ }\bibfield  {title}
  {\enquote {\bibinfo {title} {Effect of the magnon dispersion on the
  longitudinal spin seebeck effect in yttrium iron garnets},}\ }\href@noop {}
  {\bibfield  {journal} {\bibinfo  {journal} {Physical Review B}\ }\textbf
  {\bibinfo {volume} {92}},\ \bibinfo {pages} {054436} (\bibinfo {year}
  {2015})}\BibitemShut {NoStop}%
\bibitem [{\citenamefont {Ramos}\ \emph {et~al.}(2013)\citenamefont {Ramos},
  \citenamefont {Kikkawa}, \citenamefont {Uchida}, \citenamefont {Adachi},
  \citenamefont {Lucas}, \citenamefont {Aguirre}, \citenamefont {Algarabel},
  \citenamefont {Morell{\'o}n}, \citenamefont {Maekawa}, \citenamefont {Saitoh}
  \emph {et~al.}}]{ramos2013observation}%
  \BibitemOpen
  \bibfield  {author} {\bibinfo {author} {\bibfnamefont {R.}~\bibnamefont
  {Ramos}}, \bibinfo {author} {\bibfnamefont {T.}~\bibnamefont {Kikkawa}},
  \bibinfo {author} {\bibfnamefont {K.}~\bibnamefont {Uchida}}, \bibinfo
  {author} {\bibfnamefont {H.}~\bibnamefont {Adachi}}, \bibinfo {author}
  {\bibfnamefont {I.}~\bibnamefont {Lucas}}, \bibinfo {author} {\bibfnamefont
  {M.}~\bibnamefont {Aguirre}}, \bibinfo {author} {\bibfnamefont
  {P.}~\bibnamefont {Algarabel}}, \bibinfo {author} {\bibfnamefont
  {L.}~\bibnamefont {Morell{\'o}n}}, \bibinfo {author} {\bibfnamefont
  {S.}~\bibnamefont {Maekawa}}, \bibinfo {author} {\bibfnamefont
  {E.}~\bibnamefont {Saitoh}},  \emph {et~al.},\ }\bibfield  {title} {\enquote
  {\bibinfo {title} {Observation of the spin seebeck effect in epitaxial
  {F}e$_3${O}$_4$ thin films},}\ }\href@noop {} {\bibfield  {journal} {\bibinfo
   {journal} {Applied Physics Letters}\ }\textbf {\bibinfo {volume} {102}},\
  \bibinfo {pages} {072413} (\bibinfo {year} {2013})}\BibitemShut {NoStop}%
\bibitem [{\citenamefont {De}\ \emph {et~al.}(2019)\citenamefont {De},
  \citenamefont {Ghosh}, \citenamefont {Mandal}, \citenamefont {Ogale},\ and\
  \citenamefont {Nair}}]{de2019temperature}%
  \BibitemOpen
  \bibfield  {author} {\bibinfo {author} {\bibfnamefont {A.}~\bibnamefont
  {De}}, \bibinfo {author} {\bibfnamefont {A.}~\bibnamefont {Ghosh}}, \bibinfo
  {author} {\bibfnamefont {R.}~\bibnamefont {Mandal}}, \bibinfo {author}
  {\bibfnamefont {S.}~\bibnamefont {Ogale}}, \ and\ \bibinfo {author}
  {\bibfnamefont {S.}~\bibnamefont {Nair}},\ }\bibfield  {title} {\enquote
  {\bibinfo {title} {Temperature dependence of the spin seebeck effect in a
  mixed valent manganite},}\ }\href@noop {} {\bibfield  {journal} {\bibinfo
  {journal} {arXiv preprint arXiv:1905.02527}\ } (\bibinfo {year}
  {2019})}\BibitemShut {NoStop}%
\bibitem [{\citenamefont {Lin}\ \emph {et~al.}(2016)\citenamefont {Lin},
  \citenamefont {Chen}, \citenamefont {Zhang},\ and\ \citenamefont
  {Chien}}]{lin2016enhancement}%
  \BibitemOpen
  \bibfield  {author} {\bibinfo {author} {\bibfnamefont {W.}~\bibnamefont
  {Lin}}, \bibinfo {author} {\bibfnamefont {K.}~\bibnamefont {Chen}}, \bibinfo
  {author} {\bibfnamefont {S.}~\bibnamefont {Zhang}}, \ and\ \bibinfo {author}
  {\bibfnamefont {C.}~\bibnamefont {Chien}},\ }\bibfield  {title} {\enquote
  {\bibinfo {title} {Enhancement of thermally injected spin current through an
  antiferromagnetic insulator},}\ }\href@noop {} {\bibfield  {journal}
  {\bibinfo  {journal} {Physical review letters}\ }\textbf {\bibinfo {volume}
  {116}},\ \bibinfo {pages} {186601} (\bibinfo {year} {2016})}\BibitemShut
  {NoStop}%
\bibitem [{\citenamefont {Mallick}, \citenamefont {Wagh},\ and\ \citenamefont
  {Kumar}(2019)}]{mallick2019enhanced}%
  \BibitemOpen
  \bibfield  {author} {\bibinfo {author} {\bibfnamefont {K.}~\bibnamefont
  {Mallick}}, \bibinfo {author} {\bibfnamefont {A.~A.}\ \bibnamefont {Wagh}}, \
  and\ \bibinfo {author} {\bibfnamefont {P.~A.}\ \bibnamefont {Kumar}},\
  }\bibfield  {title} {\enquote {\bibinfo {title} {Enhanced spin transport in a
  ferrite having distributed energy barriers for exchange bias},}\ }\href@noop
  {} {\bibfield  {journal} {\bibinfo  {journal} {Journal of Magnetism and
  Magnetic Materials}\ }\textbf {\bibinfo {volume} {492}},\ \bibinfo {pages}
  {165644} (\bibinfo {year} {2019})}\BibitemShut {NoStop}%
\bibitem [{\citenamefont {Prakash}\ \emph {et~al.}(2016)\citenamefont
  {Prakash}, \citenamefont {Brangham}, \citenamefont {Yang},\ and\
  \citenamefont {Heremans}}]{prakash2016spin}%
  \BibitemOpen
  \bibfield  {author} {\bibinfo {author} {\bibfnamefont {A.}~\bibnamefont
  {Prakash}}, \bibinfo {author} {\bibfnamefont {J.}~\bibnamefont {Brangham}},
  \bibinfo {author} {\bibfnamefont {F.}~\bibnamefont {Yang}}, \ and\ \bibinfo
  {author} {\bibfnamefont {J.~P.}\ \bibnamefont {Heremans}},\ }\bibfield
  {title} {\enquote {\bibinfo {title} {Spin seebeck effect through
  antiferromagnetic {NiO}},}\ }\href@noop {} {\bibfield  {journal} {\bibinfo
  {journal} {Physical Review B}\ }\textbf {\bibinfo {volume} {94}},\ \bibinfo
  {pages} {014427} (\bibinfo {year} {2016})}\BibitemShut {NoStop}%
\bibitem [{\citenamefont {Qiu}\ \emph {et~al.}(2018)\citenamefont {Qiu},
  \citenamefont {Hou}, \citenamefont {Barker}, \citenamefont {Yamamoto},
  \citenamefont {Gomonay},\ and\ \citenamefont {Saitoh}}]{qiu2018spin}%
  \BibitemOpen
  \bibfield  {author} {\bibinfo {author} {\bibfnamefont {Z.}~\bibnamefont
  {Qiu}}, \bibinfo {author} {\bibfnamefont {D.}~\bibnamefont {Hou}}, \bibinfo
  {author} {\bibfnamefont {J.}~\bibnamefont {Barker}}, \bibinfo {author}
  {\bibfnamefont {K.}~\bibnamefont {Yamamoto}}, \bibinfo {author}
  {\bibfnamefont {O.}~\bibnamefont {Gomonay}}, \ and\ \bibinfo {author}
  {\bibfnamefont {E.}~\bibnamefont {Saitoh}},\ }\bibfield  {title} {\enquote
  {\bibinfo {title} {Spin colossal magnetoresistance in an antiferromagnetic
  insulator},}\ }\href@noop {} {\bibfield  {journal} {\bibinfo  {journal}
  {Nature materials}\ ,\ \bibinfo {pages} {1}} (\bibinfo {year}
  {2018})}\BibitemShut {NoStop}%
\bibitem [{\citenamefont {Cramer}\ \emph {et~al.}(2018)\citenamefont {Cramer},
  \citenamefont {Ritzmann}, \citenamefont {Dong}, \citenamefont {Jaiswal},
  \citenamefont {Qiu}, \citenamefont {Saitoh}, \citenamefont {Nowak},\ and\
  \citenamefont {Kl{\"a}ui}}]{cramer2018spin}%
  \BibitemOpen
  \bibfield  {author} {\bibinfo {author} {\bibfnamefont {J.}~\bibnamefont
  {Cramer}}, \bibinfo {author} {\bibfnamefont {U.}~\bibnamefont {Ritzmann}},
  \bibinfo {author} {\bibfnamefont {B.-W.}\ \bibnamefont {Dong}}, \bibinfo
  {author} {\bibfnamefont {S.}~\bibnamefont {Jaiswal}}, \bibinfo {author}
  {\bibfnamefont {Z.}~\bibnamefont {Qiu}}, \bibinfo {author} {\bibfnamefont
  {E.}~\bibnamefont {Saitoh}}, \bibinfo {author} {\bibfnamefont
  {U.}~\bibnamefont {Nowak}}, \ and\ \bibinfo {author} {\bibfnamefont
  {M.}~\bibnamefont {Kl{\"a}ui}},\ }\bibfield  {title} {\enquote {\bibinfo
  {title} {Spin transport across antiferromagnets induced by the spin seebeck
  effect},}\ }\href@noop {} {\bibfield  {journal} {\bibinfo  {journal} {Journal
  of Physics D: Applied Physics}\ }\textbf {\bibinfo {volume} {51}},\ \bibinfo
  {pages} {144004} (\bibinfo {year} {2018})}\BibitemShut {NoStop}%
\bibitem [{\citenamefont {Baldrati}\ \emph {et~al.}(2018)\citenamefont
  {Baldrati}, \citenamefont {Schneider}, \citenamefont {Niizeki}, \citenamefont
  {Ramos}, \citenamefont {Cramer}, \citenamefont {Ross}, \citenamefont
  {Saitoh},\ and\ \citenamefont {Kl{\"a}ui}}]{baldrati2018spin}%
  \BibitemOpen
  \bibfield  {author} {\bibinfo {author} {\bibfnamefont {L.}~\bibnamefont
  {Baldrati}}, \bibinfo {author} {\bibfnamefont {C.}~\bibnamefont {Schneider}},
  \bibinfo {author} {\bibfnamefont {T.}~\bibnamefont {Niizeki}}, \bibinfo
  {author} {\bibfnamefont {R.}~\bibnamefont {Ramos}}, \bibinfo {author}
  {\bibfnamefont {J.}~\bibnamefont {Cramer}}, \bibinfo {author} {\bibfnamefont
  {A.}~\bibnamefont {Ross}}, \bibinfo {author} {\bibfnamefont {E.}~\bibnamefont
  {Saitoh}}, \ and\ \bibinfo {author} {\bibfnamefont {M.}~\bibnamefont
  {Kl{\"a}ui}},\ }\bibfield  {title} {\enquote {\bibinfo {title} {Spin
  transport in multilayer systems with fully epitaxial {NiO} thin films},}\
  }\href@noop {} {\bibfield  {journal} {\bibinfo  {journal} {Physical Review
  B}\ }\textbf {\bibinfo {volume} {98}},\ \bibinfo {pages} {014409} (\bibinfo
  {year} {2018})}\BibitemShut {NoStop}%
\bibitem [{\citenamefont {Khymyn}\ \emph {et~al.}(2016)\citenamefont {Khymyn},
  \citenamefont {Lisenkov}, \citenamefont {Tiberkevich}, \citenamefont
  {Slavin},\ and\ \citenamefont {Ivanov}}]{khymyn2016transformation}%
  \BibitemOpen
  \bibfield  {author} {\bibinfo {author} {\bibfnamefont {R.}~\bibnamefont
  {Khymyn}}, \bibinfo {author} {\bibfnamefont {I.}~\bibnamefont {Lisenkov}},
  \bibinfo {author} {\bibfnamefont {V.~S.}\ \bibnamefont {Tiberkevich}},
  \bibinfo {author} {\bibfnamefont {A.~N.}\ \bibnamefont {Slavin}}, \ and\
  \bibinfo {author} {\bibfnamefont {B.~A.}\ \bibnamefont {Ivanov}},\ }\bibfield
   {title} {\enquote {\bibinfo {title} {Transformation of spin current by
  antiferromagnetic insulators},}\ }\href@noop {} {\bibfield  {journal}
  {\bibinfo  {journal} {Physical Review B}\ }\textbf {\bibinfo {volume} {93}},\
  \bibinfo {pages} {224421} (\bibinfo {year} {2016})}\BibitemShut {NoStop}%
\bibitem [{\citenamefont {Rezende}, \citenamefont {Rodr{\'\i}guez-Su{\'a}rez},\
  and\ \citenamefont {Azevedo}(2016)}]{rezende2016diffusive}%
  \BibitemOpen
  \bibfield  {author} {\bibinfo {author} {\bibfnamefont {S.}~\bibnamefont
  {Rezende}}, \bibinfo {author} {\bibfnamefont {R.}~\bibnamefont
  {Rodr{\'\i}guez-Su{\'a}rez}}, \ and\ \bibinfo {author} {\bibfnamefont
  {A.}~\bibnamefont {Azevedo}},\ }\bibfield  {title} {\enquote {\bibinfo
  {title} {Diffusive magnonic spin transport in antiferromagnetic
  insulators},}\ }\href@noop {} {\bibfield  {journal} {\bibinfo  {journal}
  {Physical Review B}\ }\textbf {\bibinfo {volume} {93}},\ \bibinfo {pages}
  {054412} (\bibinfo {year} {2016})}\BibitemShut {NoStop}%
\bibitem [{\citenamefont {Uchida}\ \emph
  {et~al.}(2014{\natexlab{b}})\citenamefont {Uchida}, \citenamefont {Kikkawa},
  \citenamefont {Miura}, \citenamefont {Shiomi},\ and\ \citenamefont
  {Saitoh}}]{uchida2014quantitative}%
  \BibitemOpen
  \bibfield  {author} {\bibinfo {author} {\bibfnamefont {K.-i.}\ \bibnamefont
  {Uchida}}, \bibinfo {author} {\bibfnamefont {T.}~\bibnamefont {Kikkawa}},
  \bibinfo {author} {\bibfnamefont {A.}~\bibnamefont {Miura}}, \bibinfo
  {author} {\bibfnamefont {J.}~\bibnamefont {Shiomi}}, \ and\ \bibinfo {author}
  {\bibfnamefont {E.}~\bibnamefont {Saitoh}},\ }\bibfield  {title} {\enquote
  {\bibinfo {title} {Quantitative temperature dependence of longitudinal spin
  seebeck effect at high temperatures},}\ }\href@noop {} {\bibfield  {journal}
  {\bibinfo  {journal} {Physical Review X}\ }\textbf {\bibinfo {volume} {4}},\
  \bibinfo {pages} {041023} (\bibinfo {year} {2014}{\natexlab{b}})}\BibitemShut
  {NoStop}%
\bibitem [{\citenamefont {Wu}\ \emph {et~al.}(2017)\citenamefont {Wu},
  \citenamefont {Luo}, \citenamefont {Lin},\ and\ \citenamefont
  {Huang}}]{wu2017longitudinal}%
  \BibitemOpen
  \bibfield  {author} {\bibinfo {author} {\bibfnamefont {B.}~\bibnamefont
  {Wu}}, \bibinfo {author} {\bibfnamefont {G.}~\bibnamefont {Luo}}, \bibinfo
  {author} {\bibfnamefont {J.}~\bibnamefont {Lin}}, \ and\ \bibinfo {author}
  {\bibfnamefont {S.}~\bibnamefont {Huang}},\ }\bibfield  {title} {\enquote
  {\bibinfo {title} {Longitudinal spin seebeck effect in a half-metallic
  {La}$_{0.7}${Sr}$_{0.3}${MnO}$_ 3$ film},}\ }\href@noop {} {\bibfield
  {journal} {\bibinfo  {journal} {Physical Review B}\ }\textbf {\bibinfo
  {volume} {96}},\ \bibinfo {pages} {060402} (\bibinfo {year}
  {2017})}\BibitemShut {NoStop}%
\bibitem [{\citenamefont {Adachi}\ \emph {et~al.}(2013)\citenamefont {Adachi},
  \citenamefont {Uchida}, \citenamefont {Saitoh},\ and\ \citenamefont
  {Maekawa}}]{adachi2013theory}%
  \BibitemOpen
  \bibfield  {author} {\bibinfo {author} {\bibfnamefont {H.}~\bibnamefont
  {Adachi}}, \bibinfo {author} {\bibfnamefont {K.-i.}\ \bibnamefont {Uchida}},
  \bibinfo {author} {\bibfnamefont {E.}~\bibnamefont {Saitoh}}, \ and\ \bibinfo
  {author} {\bibfnamefont {S.}~\bibnamefont {Maekawa}},\ }\bibfield  {title}
  {\enquote {\bibinfo {title} {Theory of the spin seebeck effect},}\
  }\href@noop {} {\bibfield  {journal} {\bibinfo  {journal} {Reports on
  Progress in Physics}\ }\textbf {\bibinfo {volume} {76}},\ \bibinfo {pages}
  {036501} (\bibinfo {year} {2013})}\BibitemShut {NoStop}%
\bibitem [{\citenamefont {Ohnuma}\ \emph {et~al.}(2014)\citenamefont {Ohnuma},
  \citenamefont {Adachi}, \citenamefont {Saitoh},\ and\ \citenamefont
  {Maekawa}}]{ohnuma2014enhanced}%
  \BibitemOpen
  \bibfield  {author} {\bibinfo {author} {\bibfnamefont {Y.}~\bibnamefont
  {Ohnuma}}, \bibinfo {author} {\bibfnamefont {H.}~\bibnamefont {Adachi}},
  \bibinfo {author} {\bibfnamefont {E.}~\bibnamefont {Saitoh}}, \ and\ \bibinfo
  {author} {\bibfnamefont {S.}~\bibnamefont {Maekawa}},\ }\bibfield  {title}
  {\enquote {\bibinfo {title} {Enhanced dc spin pumping into a fluctuating
  ferromagnet near {T$_C$}},}\ }\href@noop {} {\bibfield  {journal} {\bibinfo
  {journal} {Physical Review B}\ }\textbf {\bibinfo {volume} {89}},\ \bibinfo
  {pages} {174417} (\bibinfo {year} {2014})}\BibitemShut {NoStop}%
\bibitem [{\citenamefont {Adachi}, \citenamefont {Yamamoto},\ and\
  \citenamefont {Ichioka}(2018)}]{adachi2018spin}%
  \BibitemOpen
  \bibfield  {author} {\bibinfo {author} {\bibfnamefont {H.}~\bibnamefont
  {Adachi}}, \bibinfo {author} {\bibfnamefont {Y.}~\bibnamefont {Yamamoto}}, \
  and\ \bibinfo {author} {\bibfnamefont {M.}~\bibnamefont {Ichioka}},\
  }\bibfield  {title} {\enquote {\bibinfo {title} {Spin seebeck effect in a
  simple ferromagnet near {T$_C$} : a ginzburg--landau approach},}\ }\href@noop
  {} {\bibfield  {journal} {\bibinfo  {journal} {Journal of Physics D: Applied
  Physics}\ }\textbf {\bibinfo {volume} {51}},\ \bibinfo {pages} {144001}
  (\bibinfo {year} {2018})}\BibitemShut {NoStop}%
\bibitem [{\citenamefont {Barker}\ and\ \citenamefont
  {Bauer}(2016)}]{barker2016thermal}%
  \BibitemOpen
  \bibfield  {author} {\bibinfo {author} {\bibfnamefont {J.}~\bibnamefont
  {Barker}}\ and\ \bibinfo {author} {\bibfnamefont {G.~E.}\ \bibnamefont
  {Bauer}},\ }\bibfield  {title} {\enquote {\bibinfo {title} {Thermal spin
  dynamics of yttrium iron garnet},}\ }\href@noop {} {\bibfield  {journal}
  {\bibinfo  {journal} {Physical review letters}\ }\textbf {\bibinfo {volume}
  {117}},\ \bibinfo {pages} {217201} (\bibinfo {year} {2016})}\BibitemShut
  {NoStop}%
\bibitem [{\citenamefont {Levy}(1969)}]{levy1969spontaneous}%
  \BibitemOpen
  \bibfield  {author} {\bibinfo {author} {\bibfnamefont {F.}~\bibnamefont
  {Levy}},\ }\bibfield  {title} {\enquote {\bibinfo {title} {Spontaneous
  magnetoelastic effects in some rare earth compounds. i.
  europium-chalcogenides},}\ }\href@noop {} {\bibfield  {journal} {\bibinfo
  {journal} {PHYSIK DER KONDENSITERTEN MATERIE}\ }\textbf {\bibinfo {volume}
  {10}},\ \bibinfo {pages} {71--+} (\bibinfo {year} {1969})}\BibitemShut
  {NoStop}%
\bibitem [{\citenamefont {Dietrich}, \citenamefont {Henderson~Jr},\ and\
  \citenamefont {Meyer}(1975)}]{dietrich1975spin}%
  \BibitemOpen
  \bibfield  {author} {\bibinfo {author} {\bibfnamefont {O.}~\bibnamefont
  {Dietrich}}, \bibinfo {author} {\bibfnamefont {A.}~\bibnamefont
  {Henderson~Jr}}, \ and\ \bibinfo {author} {\bibfnamefont {H.}~\bibnamefont
  {Meyer}},\ }\bibfield  {title} {\enquote {\bibinfo {title} {Spin-wave
  analysis of specific heat and magnetization in {EuO} and {EuS}},}\
  }\href@noop {} {\bibfield  {journal} {\bibinfo  {journal} {Physical Review
  B}\ }\textbf {\bibinfo {volume} {12}},\ \bibinfo {pages} {2844} (\bibinfo
  {year} {1975})}\BibitemShut {NoStop}%
\bibitem [{\citenamefont {Kasuya}(1970)}]{kasuya1970exchange}%
  \BibitemOpen
  \bibfield  {author} {\bibinfo {author} {\bibfnamefont {T.}~\bibnamefont
  {Kasuya}},\ }\bibfield  {title} {\enquote {\bibinfo {title} {Exchange
  mechanisms in europium chalcogenides},}\ }\href@noop {} {\bibfield  {journal}
  {\bibinfo  {journal} {IBM Journal of Research and Development}\ }\textbf
  {\bibinfo {volume} {14}},\ \bibinfo {pages} {214--223} (\bibinfo {year}
  {1970})}\BibitemShut {NoStop}%
\bibitem [{\citenamefont {Passell}, \citenamefont {Dietrich},\ and\
  \citenamefont {Als-Nielsen}(1976)}]{passell1976neutron}%
  \BibitemOpen
  \bibfield  {author} {\bibinfo {author} {\bibfnamefont {L.}~\bibnamefont
  {Passell}}, \bibinfo {author} {\bibfnamefont {O.}~\bibnamefont {Dietrich}}, \
  and\ \bibinfo {author} {\bibfnamefont {J.}~\bibnamefont {Als-Nielsen}},\
  }\bibfield  {title} {\enquote {\bibinfo {title} {Neutron scattering from the
  heisenberg ferromagnets {EuO} and {EuS}. i. the exchange interactions},}\
  }\href@noop {} {\bibfield  {journal} {\bibinfo  {journal} {Physical Review
  B}\ }\textbf {\bibinfo {volume} {14}},\ \bibinfo {pages} {4897} (\bibinfo
  {year} {1976})}\BibitemShut {NoStop}%
\bibitem [{\citenamefont {Oliver}\ \emph {et~al.}(1972)\citenamefont {Oliver},
  \citenamefont {Dimmock}, \citenamefont {McWhorter},\ and\ \citenamefont
  {Reed}}]{oliver1972conductivity}%
  \BibitemOpen
  \bibfield  {author} {\bibinfo {author} {\bibfnamefont {M.~R.}\ \bibnamefont
  {Oliver}}, \bibinfo {author} {\bibfnamefont {J.}~\bibnamefont {Dimmock}},
  \bibinfo {author} {\bibfnamefont {A.}~\bibnamefont {McWhorter}}, \ and\
  \bibinfo {author} {\bibfnamefont {T.}~\bibnamefont {Reed}},\ }\bibfield
  {title} {\enquote {\bibinfo {title} {Conductivity studies in europium
  oxide},}\ }\href@noop {} {\bibfield  {journal} {\bibinfo  {journal} {Physical
  Review B}\ }\textbf {\bibinfo {volume} {5}},\ \bibinfo {pages} {1078}
  (\bibinfo {year} {1972})}\BibitemShut {NoStop}%
\bibitem [{\citenamefont {Penney}, \citenamefont {Shafer},\ and\ \citenamefont
  {Torrance}(1972)}]{penney1972insulator}%
  \BibitemOpen
  \bibfield  {author} {\bibinfo {author} {\bibfnamefont {T.}~\bibnamefont
  {Penney}}, \bibinfo {author} {\bibfnamefont {M.}~\bibnamefont {Shafer}}, \
  and\ \bibinfo {author} {\bibfnamefont {J.}~\bibnamefont {Torrance}},\
  }\bibfield  {title} {\enquote {\bibinfo {title} {Insulator-metal transition
  and long-range magnetic order in {EuO}},}\ }\href@noop {} {\bibfield
  {journal} {\bibinfo  {journal} {Physical Review B}\ }\textbf {\bibinfo
  {volume} {5}},\ \bibinfo {pages} {3669} (\bibinfo {year} {1972})}\BibitemShut
  {NoStop}%
\bibitem [{\citenamefont {Steeneken}\ \emph {et~al.}(2002)\citenamefont
  {Steeneken}, \citenamefont {Tjeng}, \citenamefont {Elfimov}, \citenamefont
  {Sawatzky}, \citenamefont {Ghiringhelli}, \citenamefont {Brookes},\ and\
  \citenamefont {Huang}}]{steeneken2002exchange}%
  \BibitemOpen
  \bibfield  {author} {\bibinfo {author} {\bibfnamefont {P.}~\bibnamefont
  {Steeneken}}, \bibinfo {author} {\bibfnamefont {L.}~\bibnamefont {Tjeng}},
  \bibinfo {author} {\bibfnamefont {I.}~\bibnamefont {Elfimov}}, \bibinfo
  {author} {\bibfnamefont {G.}~\bibnamefont {Sawatzky}}, \bibinfo {author}
  {\bibfnamefont {G.}~\bibnamefont {Ghiringhelli}}, \bibinfo {author}
  {\bibfnamefont {N.}~\bibnamefont {Brookes}}, \ and\ \bibinfo {author}
  {\bibfnamefont {D.-J.}\ \bibnamefont {Huang}},\ }\bibfield  {title} {\enquote
  {\bibinfo {title} {Exchange splitting and charge carrier spin polarization in
  {EuO}},}\ }\href@noop {} {\bibfield  {journal} {\bibinfo  {journal} {Physical
  review letters}\ }\textbf {\bibinfo {volume} {88}},\ \bibinfo {pages}
  {047201} (\bibinfo {year} {2002})}\BibitemShut {NoStop}%
\bibitem [{\citenamefont {Schmehl}\ \emph {et~al.}(2007)\citenamefont
  {Schmehl}, \citenamefont {Vaithyanathan}, \citenamefont {Herrnberger},
  \citenamefont {Thiel}, \citenamefont {Richter}, \citenamefont {Liberati},
  \citenamefont {Heeg}, \citenamefont {R{\"o}ckerath}, \citenamefont
  {Kourkoutis}, \citenamefont {M{\"u}hlbauer} \emph
  {et~al.}}]{schmehl2007epitaxial}%
  \BibitemOpen
  \bibfield  {author} {\bibinfo {author} {\bibfnamefont {A.}~\bibnamefont
  {Schmehl}}, \bibinfo {author} {\bibfnamefont {V.}~\bibnamefont
  {Vaithyanathan}}, \bibinfo {author} {\bibfnamefont {A.}~\bibnamefont
  {Herrnberger}}, \bibinfo {author} {\bibfnamefont {S.}~\bibnamefont {Thiel}},
  \bibinfo {author} {\bibfnamefont {C.}~\bibnamefont {Richter}}, \bibinfo
  {author} {\bibfnamefont {M.}~\bibnamefont {Liberati}}, \bibinfo {author}
  {\bibfnamefont {T.}~\bibnamefont {Heeg}}, \bibinfo {author} {\bibfnamefont
  {M.}~\bibnamefont {R{\"o}ckerath}}, \bibinfo {author} {\bibfnamefont {L.~F.}\
  \bibnamefont {Kourkoutis}}, \bibinfo {author} {\bibfnamefont
  {S.}~\bibnamefont {M{\"u}hlbauer}},  \emph {et~al.},\ }\bibfield  {title}
  {\enquote {\bibinfo {title} {Epitaxial integration of the highly
  spin-polarized ferromagnetic semiconductor {EuO} with silicon and {GaN}},}\
  }\href@noop {} {\bibfield  {journal} {\bibinfo  {journal} {Nature materials}\
  }\textbf {\bibinfo {volume} {6}},\ \bibinfo {pages} {882} (\bibinfo {year}
  {2007})}\BibitemShut {NoStop}%
\bibitem [{\citenamefont {Barbagallo}\ \emph {et~al.}(2010)\citenamefont
  {Barbagallo}, \citenamefont {Hine}, \citenamefont {Cooper}, \citenamefont
  {Steinke}, \citenamefont {Ionescu}, \citenamefont {Barnes}, \citenamefont
  {Kinane}, \citenamefont {Dalgliesh}, \citenamefont {Charlton},\ and\
  \citenamefont {Langridge}}]{barbagallo2010experimental}%
  \BibitemOpen
  \bibfield  {author} {\bibinfo {author} {\bibfnamefont {M.}~\bibnamefont
  {Barbagallo}}, \bibinfo {author} {\bibfnamefont {N.}~\bibnamefont {Hine}},
  \bibinfo {author} {\bibfnamefont {J.}~\bibnamefont {Cooper}}, \bibinfo
  {author} {\bibfnamefont {N.-J.}\ \bibnamefont {Steinke}}, \bibinfo {author}
  {\bibfnamefont {A.}~\bibnamefont {Ionescu}}, \bibinfo {author} {\bibfnamefont
  {C.}~\bibnamefont {Barnes}}, \bibinfo {author} {\bibfnamefont
  {C.}~\bibnamefont {Kinane}}, \bibinfo {author} {\bibfnamefont
  {R.}~\bibnamefont {Dalgliesh}}, \bibinfo {author} {\bibfnamefont
  {T.}~\bibnamefont {Charlton}}, \ and\ \bibinfo {author} {\bibfnamefont
  {S.}~\bibnamefont {Langridge}},\ }\bibfield  {title} {\enquote {\bibinfo
  {title} {Experimental and theoretical analysis of magnetic moment enhancement
  in oxygen-deficient {E}u{O}},}\ }\href@noop {} {\bibfield  {journal}
  {\bibinfo  {journal} {Physical Review B}\ }\textbf {\bibinfo {volume} {81}},\
  \bibinfo {pages} {235216} (\bibinfo {year} {2010})}\BibitemShut {NoStop}%
\bibitem [{\citenamefont {Jansen}(2012)}]{jansen2012silicon}%
  \BibitemOpen
  \bibfield  {author} {\bibinfo {author} {\bibfnamefont {R.}~\bibnamefont
  {Jansen}},\ }\bibfield  {title} {\enquote {\bibinfo {title} {Silicon
  spintronics},}\ }\href@noop {} {\bibfield  {journal} {\bibinfo  {journal}
  {Nature materials}\ }\textbf {\bibinfo {volume} {11}},\ \bibinfo {pages}
  {400} (\bibinfo {year} {2012})}\BibitemShut {NoStop}%
\bibitem [{\citenamefont {Dass}, \citenamefont {Yan},\ and\ \citenamefont
  {Goodenough}(2003)}]{dass2003oxygen}%
  \BibitemOpen
  \bibfield  {author} {\bibinfo {author} {\bibfnamefont {R.}~\bibnamefont
  {Dass}}, \bibinfo {author} {\bibfnamefont {J.-Q.}\ \bibnamefont {Yan}}, \
  and\ \bibinfo {author} {\bibfnamefont {J.}~\bibnamefont {Goodenough}},\
  }\bibfield  {title} {\enquote {\bibinfo {title} {Oxygen stoichiometry,
  ferromagnetism, and transport properties of {L}a$_{2-x}${NiMnO}6+$\delta$},}\
  }\href@noop {} {\bibfield  {journal} {\bibinfo  {journal} {Physical Review
  B}\ }\textbf {\bibinfo {volume} {68}},\ \bibinfo {pages} {064415} (\bibinfo
  {year} {2003})}\BibitemShut {NoStop}%
\bibitem [{\citenamefont {Das}\ \emph {et~al.}(2008)\citenamefont {Das},
  \citenamefont {Waghmare}, \citenamefont {Saha-Dasgupta},\ and\ \citenamefont
  {Sarma}}]{das2008electronic}%
  \BibitemOpen
  \bibfield  {author} {\bibinfo {author} {\bibfnamefont {H.}~\bibnamefont
  {Das}}, \bibinfo {author} {\bibfnamefont {U.~V.}\ \bibnamefont {Waghmare}},
  \bibinfo {author} {\bibfnamefont {T.}~\bibnamefont {Saha-Dasgupta}}, \ and\
  \bibinfo {author} {\bibfnamefont {D.}~\bibnamefont {Sarma}},\ }\bibfield
  {title} {\enquote {\bibinfo {title} {Electronic structure, phonons, and
  dielectric anomaly in ferromagnetic insulating double pervoskite
  {L}a$_2${N}i{M}no$_6$},}\ }\href@noop {} {\bibfield  {journal} {\bibinfo
  {journal} {Physical review letters}\ }\textbf {\bibinfo {volume} {100}},\
  \bibinfo {pages} {186402} (\bibinfo {year} {2008})}\BibitemShut {NoStop}%
\bibitem [{\citenamefont {Kumar}\ \emph {et~al.}(2014)\citenamefont {Kumar},
  \citenamefont {Ghara}, \citenamefont {Rajeswaran}, \citenamefont {Muthu},
  \citenamefont {Sundaresan},\ and\ \citenamefont
  {Sood}}]{kumar2014temperature}%
  \BibitemOpen
  \bibfield  {author} {\bibinfo {author} {\bibfnamefont {P.}~\bibnamefont
  {Kumar}}, \bibinfo {author} {\bibfnamefont {S.}~\bibnamefont {Ghara}},
  \bibinfo {author} {\bibfnamefont {B.}~\bibnamefont {Rajeswaran}}, \bibinfo
  {author} {\bibfnamefont {D.}~\bibnamefont {Muthu}}, \bibinfo {author}
  {\bibfnamefont {A.}~\bibnamefont {Sundaresan}}, \ and\ \bibinfo {author}
  {\bibfnamefont {A.}~\bibnamefont {Sood}},\ }\bibfield  {title} {\enquote
  {\bibinfo {title} {Temperature dependent magnetic, dielectric and raman
  studies of partially disordered {L}a$_2${N}i{M}no$_6$},}\ }\href@noop {}
  {\bibfield  {journal} {\bibinfo  {journal} {Solid State Communications}\
  }\textbf {\bibinfo {volume} {184}},\ \bibinfo {pages} {47--51} (\bibinfo
  {year} {2014})}\BibitemShut {NoStop}%
\bibitem [{\citenamefont {Guo}\ \emph {et~al.}(2006)\citenamefont {Guo},
  \citenamefont {Burgess}, \citenamefont {Street}, \citenamefont {Gupta},
  \citenamefont {Calvarese},\ and\ \citenamefont
  {Subramanian}}]{guo2006growth}%
  \BibitemOpen
  \bibfield  {author} {\bibinfo {author} {\bibfnamefont {H.}~\bibnamefont
  {Guo}}, \bibinfo {author} {\bibfnamefont {J.}~\bibnamefont {Burgess}},
  \bibinfo {author} {\bibfnamefont {S.}~\bibnamefont {Street}}, \bibinfo
  {author} {\bibfnamefont {A.}~\bibnamefont {Gupta}}, \bibinfo {author}
  {\bibfnamefont {T.}~\bibnamefont {Calvarese}}, \ and\ \bibinfo {author}
  {\bibfnamefont {M.}~\bibnamefont {Subramanian}},\ }\bibfield  {title}
  {\enquote {\bibinfo {title} {Growth of epitaxial thin films of the ordered
  double perovskite {L}a$_2${N}i{M}no$_6$ on different substrates},}\
  }\href@noop {} {\bibfield  {journal} {\bibinfo  {journal} {Applied physics
  letters}\ }\textbf {\bibinfo {volume} {89}},\ \bibinfo {pages} {022509}
  (\bibinfo {year} {2006})}\BibitemShut {NoStop}%
\bibitem [{\citenamefont {Shiomi}\ and\ \citenamefont
  {Saitoh}(2014)}]{shiomi2014paramagnetic}%
  \BibitemOpen
  \bibfield  {author} {\bibinfo {author} {\bibfnamefont {Y.}~\bibnamefont
  {Shiomi}}\ and\ \bibinfo {author} {\bibfnamefont {E.}~\bibnamefont
  {Saitoh}},\ }\bibfield  {title} {\enquote {\bibinfo {title} {Paramagnetic
  spin pumping},}\ }\href@noop {} {\bibfield  {journal} {\bibinfo  {journal}
  {Physical review letters}\ }\textbf {\bibinfo {volume} {113}},\ \bibinfo
  {pages} {266602} (\bibinfo {year} {2014})}\BibitemShut {NoStop}%
\bibitem [{\citenamefont {Hashisaka}\ \emph {et~al.}(2007)\citenamefont
  {Hashisaka}, \citenamefont {Kan}, \citenamefont {Masuno}, \citenamefont
  {Terashima}, \citenamefont {Takano},\ and\ \citenamefont
  {Mibu}}]{hashisaka2007spin}%
  \BibitemOpen
  \bibfield  {author} {\bibinfo {author} {\bibfnamefont {M.}~\bibnamefont
  {Hashisaka}}, \bibinfo {author} {\bibfnamefont {D.}~\bibnamefont {Kan}},
  \bibinfo {author} {\bibfnamefont {A.}~\bibnamefont {Masuno}}, \bibinfo
  {author} {\bibfnamefont {T.}~\bibnamefont {Terashima}}, \bibinfo {author}
  {\bibfnamefont {M.}~\bibnamefont {Takano}}, \ and\ \bibinfo {author}
  {\bibfnamefont {K.}~\bibnamefont {Mibu}},\ }\bibfield  {title} {\enquote
  {\bibinfo {title} {Spin-filtering effect of ferromagnetic semiconductor
  {L}a$_2${N}i{M}no$_6$},}\ }\href@noop {} {\bibfield  {journal} {\bibinfo
  {journal} {Journal of Magnetism and Magnetic Materials}\ }\textbf {\bibinfo
  {volume} {310}},\ \bibinfo {pages} {1975--1977} (\bibinfo {year}
  {2007})}\BibitemShut {NoStop}%
\bibitem [{\citenamefont {Zhou}\ \emph {et~al.}(2007)\citenamefont {Zhou},
  \citenamefont {Shi}, \citenamefont {Yang},\ and\ \citenamefont
  {Zhao}}]{zhou2007evidence}%
  \BibitemOpen
  \bibfield  {author} {\bibinfo {author} {\bibfnamefont {S.}~\bibnamefont
  {Zhou}}, \bibinfo {author} {\bibfnamefont {L.}~\bibnamefont {Shi}}, \bibinfo
  {author} {\bibfnamefont {H.}~\bibnamefont {Yang}}, \ and\ \bibinfo {author}
  {\bibfnamefont {J.}~\bibnamefont {Zhao}},\ }\bibfield  {title} {\enquote
  {\bibinfo {title} {Evidence of short-range magnetic ordering above tc in the
  double perovskite {L}a$_2${N}i{M}no$_6$},}\ }\href@noop {} {\bibfield
  {journal} {\bibinfo  {journal} {Applied Physics Letters}\ }\textbf {\bibinfo
  {volume} {91}},\ \bibinfo {pages} {172505} (\bibinfo {year}
  {2007})}\BibitemShut {NoStop}%
\bibitem [{\citenamefont {Guo}\ \emph {et~al.}(2009)\citenamefont {Guo},
  \citenamefont {Gupta}, \citenamefont {Varela}, \citenamefont {Pennycook},\
  and\ \citenamefont {Zhang}}]{guo2009local}%
  \BibitemOpen
  \bibfield  {author} {\bibinfo {author} {\bibfnamefont {H.}~\bibnamefont
  {Guo}}, \bibinfo {author} {\bibfnamefont {A.}~\bibnamefont {Gupta}}, \bibinfo
  {author} {\bibfnamefont {M.}~\bibnamefont {Varela}}, \bibinfo {author}
  {\bibfnamefont {S.}~\bibnamefont {Pennycook}}, \ and\ \bibinfo {author}
  {\bibfnamefont {J.}~\bibnamefont {Zhang}},\ }\bibfield  {title} {\enquote
  {\bibinfo {title} {Local valence and magnetic characteristics of
  {L}a$_2${N}i{M}no$_6$},}\ }\href@noop {} {\bibfield  {journal} {\bibinfo
  {journal} {Physical Review B}\ }\textbf {\bibinfo {volume} {79}},\ \bibinfo
  {pages} {172402} (\bibinfo {year} {2009})}\BibitemShut {NoStop}%
\bibitem [{\citenamefont {Bull}, \citenamefont {Gleeson},\ and\ \citenamefont
  {Knight}(2003)}]{bull2003determination}%
  \BibitemOpen
  \bibfield  {author} {\bibinfo {author} {\bibfnamefont {C.}~\bibnamefont
  {Bull}}, \bibinfo {author} {\bibfnamefont {D.}~\bibnamefont {Gleeson}}, \
  and\ \bibinfo {author} {\bibfnamefont {K.}~\bibnamefont {Knight}},\
  }\bibfield  {title} {\enquote {\bibinfo {title} {Determination of b-site
  ordering and structural transformations in the mixed transition metal
  perovskites {L}a$_2${C}o{M}no$_6$ and {L}a$_2${N}i{M}no$_6$},}\ }\href@noop
  {} {\bibfield  {journal} {\bibinfo  {journal} {Journal of Physics: Condensed
  Matter}\ }\textbf {\bibinfo {volume} {15}},\ \bibinfo {pages} {4927}
  (\bibinfo {year} {2003})}\BibitemShut {NoStop}%
\bibitem [{\citenamefont {Sakurai}\ \emph {et~al.}(2011)\citenamefont
  {Sakurai}, \citenamefont {Ohkubo}, \citenamefont {Matsumoto}, \citenamefont
  {Koinuma},\ and\ \citenamefont {Oshima}}]{sakurai2011influence}%
  \BibitemOpen
  \bibfield  {author} {\bibinfo {author} {\bibfnamefont {Y.}~\bibnamefont
  {Sakurai}}, \bibinfo {author} {\bibfnamefont {I.}~\bibnamefont {Ohkubo}},
  \bibinfo {author} {\bibfnamefont {Y.}~\bibnamefont {Matsumoto}}, \bibinfo
  {author} {\bibfnamefont {H.}~\bibnamefont {Koinuma}}, \ and\ \bibinfo
  {author} {\bibfnamefont {M.}~\bibnamefont {Oshima}},\ }\bibfield  {title}
  {\enquote {\bibinfo {title} {Influence of substrates on epitaxial growth of
  b-site-ordered perovskite {L}a$_2${N}i{M}no$_6$ thin films},}\ }\href@noop {}
  {\bibfield  {journal} {\bibinfo  {journal} {Journal of Applied Physics}\
  }\textbf {\bibinfo {volume} {110}},\ \bibinfo {pages} {063913} (\bibinfo
  {year} {2011})}\BibitemShut {NoStop}%
\bibitem [{\citenamefont {Bernal-Salamanca}\ \emph {et~al.}(2019)\citenamefont
  {Bernal-Salamanca}, \citenamefont {Konstantinovi{\'c}}, \citenamefont
  {Balcells}, \citenamefont {Pannunzio-Miner}, \citenamefont {Sandiumenge},
  \citenamefont {L{\'o}pez-Mir}, \citenamefont {Bozzo}, \citenamefont
  {Herrero-Mart{\'\i}n}, \citenamefont {Pomar}, \citenamefont {Frontera} \emph
  {et~al.}}]{bernal2019non}%
  \BibitemOpen
  \bibfield  {author} {\bibinfo {author} {\bibfnamefont {M.}~\bibnamefont
  {Bernal-Salamanca}}, \bibinfo {author} {\bibfnamefont {Z.}~\bibnamefont
  {Konstantinovi{\'c}}}, \bibinfo {author} {\bibfnamefont {L.}~\bibnamefont
  {Balcells}}, \bibinfo {author} {\bibfnamefont {E.}~\bibnamefont
  {Pannunzio-Miner}}, \bibinfo {author} {\bibfnamefont {F.}~\bibnamefont
  {Sandiumenge}}, \bibinfo {author} {\bibfnamefont {L.}~\bibnamefont
  {L{\'o}pez-Mir}}, \bibinfo {author} {\bibfnamefont {B.}~\bibnamefont
  {Bozzo}}, \bibinfo {author} {\bibfnamefont {J.}~\bibnamefont
  {Herrero-Mart{\'\i}n}}, \bibinfo {author} {\bibfnamefont {A.}~\bibnamefont
  {Pomar}}, \bibinfo {author} {\bibfnamefont {C.}~\bibnamefont {Frontera}},
  \emph {et~al.},\ }\bibfield  {title} {\enquote {\bibinfo {title}
  {Non-stoichiometry driven ferromagnetism in double perovskite
  {L}a$_2${N}i$_{1-x}${M}n$_{1+x}$o$_6$ insulating thin films},}\ }\href@noop
  {} {\bibfield  {journal} {\bibinfo  {journal} {Crystal Growth \& Design}\ }
  (\bibinfo {year} {2019})}\BibitemShut {NoStop}%
\bibitem [{\citenamefont {Samokhvalov}\ \emph {et~al.}(1978)\citenamefont
  {Samokhvalov}, \citenamefont {Gunichev}, \citenamefont {Gizhevskii},
  \citenamefont {Loshkareva}, \citenamefont {Chebotaev},\ and\ \citenamefont
  {Viglin}}]{samokhvalov1978nonstoichiometric}%
  \BibitemOpen
  \bibfield  {author} {\bibinfo {author} {\bibfnamefont {A.}~\bibnamefont
  {Samokhvalov}}, \bibinfo {author} {\bibfnamefont {A.}~\bibnamefont
  {Gunichev}}, \bibinfo {author} {\bibfnamefont {B.}~\bibnamefont
  {Gizhevskii}}, \bibinfo {author} {\bibfnamefont {N.}~\bibnamefont
  {Loshkareva}}, \bibinfo {author} {\bibfnamefont {N.}~\bibnamefont
  {Chebotaev}}, \ and\ \bibinfo {author} {\bibfnamefont {N.}~\bibnamefont
  {Viglin}},\ }\bibfield  {title} {\enquote {\bibinfo {title}
  {Nonstoichiometric {E}u{O} films with an elevated curie temperature},}\
  }\href@noop {} {\bibfield  {journal} {\bibinfo  {journal} {Sov. Phys.-Solid
  State (Engl. Transl.);(United States)}\ }\textbf {\bibinfo {volume} {20}}
  (\bibinfo {year} {1978})}\BibitemShut {NoStop}%
\bibitem [{\citenamefont {Mauger}\ and\ \citenamefont
  {Godart}(1986)}]{mauger1986magnetic}%
  \BibitemOpen
  \bibfield  {author} {\bibinfo {author} {\bibfnamefont {A.}~\bibnamefont
  {Mauger}}\ and\ \bibinfo {author} {\bibfnamefont {C.}~\bibnamefont
  {Godart}},\ }\bibfield  {title} {\enquote {\bibinfo {title} {The magnetic,
  optical, and transport properties of representatives of a class of magnetic
  semiconductors: The europium chalcogenides},}\ }\href@noop {} {\bibfield
  {journal} {\bibinfo  {journal} {Physics Reports}\ }\textbf {\bibinfo {volume}
  {141}},\ \bibinfo {pages} {51--176} (\bibinfo {year} {1986})}\BibitemShut
  {NoStop}%
\bibitem [{\citenamefont {Arnold}\ and\ \citenamefont
  {Kroha}(2008)}]{arnold2008simultaneous}%
  \BibitemOpen
  \bibfield  {author} {\bibinfo {author} {\bibfnamefont {M.}~\bibnamefont
  {Arnold}}\ and\ \bibinfo {author} {\bibfnamefont {J.}~\bibnamefont {Kroha}},\
  }\bibfield  {title} {\enquote {\bibinfo {title} {Simultaneous ferromagnetic
  metal-semiconductor transition in electron-doped {E}u{O}},}\ }\href@noop {}
  {\bibfield  {journal} {\bibinfo  {journal} {Physical review letters}\
  }\textbf {\bibinfo {volume} {100}},\ \bibinfo {pages} {046404} (\bibinfo
  {year} {2008})}\BibitemShut {NoStop}%
\bibitem [{\citenamefont {Sola}\ \emph {et~al.}(2017)\citenamefont {Sola},
  \citenamefont {Bougiatioti}, \citenamefont {Kuepferling}, \citenamefont
  {Meier}, \citenamefont {Reiss}, \citenamefont {Pasquale}, \citenamefont
  {Kuschel},\ and\ \citenamefont {Basso}}]{sola2017longitudinal}%
  \BibitemOpen
  \bibfield  {author} {\bibinfo {author} {\bibfnamefont {A.}~\bibnamefont
  {Sola}}, \bibinfo {author} {\bibfnamefont {P.}~\bibnamefont {Bougiatioti}},
  \bibinfo {author} {\bibfnamefont {M.}~\bibnamefont {Kuepferling}}, \bibinfo
  {author} {\bibfnamefont {D.}~\bibnamefont {Meier}}, \bibinfo {author}
  {\bibfnamefont {G.}~\bibnamefont {Reiss}}, \bibinfo {author} {\bibfnamefont
  {M.}~\bibnamefont {Pasquale}}, \bibinfo {author} {\bibfnamefont
  {T.}~\bibnamefont {Kuschel}}, \ and\ \bibinfo {author} {\bibfnamefont
  {V.}~\bibnamefont {Basso}},\ }\bibfield  {title} {\enquote {\bibinfo {title}
  {Longitudinal spin seebeck coefficient: heat flux vs. temperature difference
  method},}\ }\href@noop {} {\bibfield  {journal} {\bibinfo  {journal}
  {Scientific reports}\ }\textbf {\bibinfo {volume} {7}},\ \bibinfo {pages}
  {46752} (\bibinfo {year} {2017})}\BibitemShut {NoStop}%
\bibitem [{\citenamefont {Prakash}\ \emph {et~al.}(2018)\citenamefont
  {Prakash}, \citenamefont {Flebus}, \citenamefont {Brangham}, \citenamefont
  {Yang}, \citenamefont {Tserkovnyak},\ and\ \citenamefont
  {Heremans}}]{prakash2018evidence}%
  \BibitemOpen
  \bibfield  {author} {\bibinfo {author} {\bibfnamefont {A.}~\bibnamefont
  {Prakash}}, \bibinfo {author} {\bibfnamefont {B.}~\bibnamefont {Flebus}},
  \bibinfo {author} {\bibfnamefont {J.}~\bibnamefont {Brangham}}, \bibinfo
  {author} {\bibfnamefont {F.}~\bibnamefont {Yang}}, \bibinfo {author}
  {\bibfnamefont {Y.}~\bibnamefont {Tserkovnyak}}, \ and\ \bibinfo {author}
  {\bibfnamefont {J.~P.}\ \bibnamefont {Heremans}},\ }\bibfield  {title}
  {\enquote {\bibinfo {title} {Evidence for the role of the magnon energy
  relaxation length in the spin seebeck effect},}\ }\href@noop {} {\bibfield
  {journal} {\bibinfo  {journal} {Physical Review B}\ }\textbf {\bibinfo
  {volume} {97}},\ \bibinfo {pages} {020408} (\bibinfo {year}
  {2018})}\BibitemShut {NoStop}%
\bibitem [{\citenamefont {Guo}\ \emph {et~al.}(2016)\citenamefont {Guo},
  \citenamefont {Cramer}, \citenamefont {Kehlberger}, \citenamefont {Ferguson},
  \citenamefont {MacLaren}, \citenamefont {Jakob},\ and\ \citenamefont
  {Kl{\"a}ui}}]{guo2016influence}%
  \BibitemOpen
  \bibfield  {author} {\bibinfo {author} {\bibfnamefont {E.-J.}\ \bibnamefont
  {Guo}}, \bibinfo {author} {\bibfnamefont {J.}~\bibnamefont {Cramer}},
  \bibinfo {author} {\bibfnamefont {A.}~\bibnamefont {Kehlberger}}, \bibinfo
  {author} {\bibfnamefont {C.~A.}\ \bibnamefont {Ferguson}}, \bibinfo {author}
  {\bibfnamefont {D.~A.}\ \bibnamefont {MacLaren}}, \bibinfo {author}
  {\bibfnamefont {G.}~\bibnamefont {Jakob}}, \ and\ \bibinfo {author}
  {\bibfnamefont {M.}~\bibnamefont {Kl{\"a}ui}},\ }\bibfield  {title} {\enquote
  {\bibinfo {title} {Influence of thickness and interface on the
  low-temperature enhancement of the spin seebeck effect in yig films},}\
  }\href@noop {} {\bibfield  {journal} {\bibinfo  {journal} {Physical Review
  X}\ }\textbf {\bibinfo {volume} {6}},\ \bibinfo {pages} {031012} (\bibinfo
  {year} {2016})}\BibitemShut {NoStop}%
\bibitem [{\citenamefont {Kalappattil}\ \emph {et~al.}(2017)\citenamefont
  {Kalappattil}, \citenamefont {Das}, \citenamefont {Phan},\ and\ \citenamefont
  {Srikanth}}]{kalappattil2017roles}%
  \BibitemOpen
  \bibfield  {author} {\bibinfo {author} {\bibfnamefont {V.}~\bibnamefont
  {Kalappattil}}, \bibinfo {author} {\bibfnamefont {R.}~\bibnamefont {Das}},
  \bibinfo {author} {\bibfnamefont {M.-H.}\ \bibnamefont {Phan}}, \ and\
  \bibinfo {author} {\bibfnamefont {H.}~\bibnamefont {Srikanth}},\ }\bibfield
  {title} {\enquote {\bibinfo {title} {Roles of bulk and surface magnetic
  anisotropy on the longitudinal spin seebeck effect of {P}t/{YIG}},}\
  }\href@noop {} {\bibfield  {journal} {\bibinfo  {journal} {Scientific
  reports}\ }\textbf {\bibinfo {volume} {7}},\ \bibinfo {pages} {13316}
  (\bibinfo {year} {2017})}\BibitemShut {NoStop}%
\bibitem [{\citenamefont {Bougiatioti}\ \emph {et~al.}(2017)\citenamefont
  {Bougiatioti}, \citenamefont {Klewe}, \citenamefont {Meier}, \citenamefont
  {Manos}, \citenamefont {Kuschel}, \citenamefont {Wollschl{\"a}ger},
  \citenamefont {Bouchenoire}, \citenamefont {Brown}, \citenamefont
  {Schmalhorst}, \citenamefont {Reiss} \emph
  {et~al.}}]{bougiatioti2017quantitative}%
  \BibitemOpen
  \bibfield  {author} {\bibinfo {author} {\bibfnamefont {P.}~\bibnamefont
  {Bougiatioti}}, \bibinfo {author} {\bibfnamefont {C.}~\bibnamefont {Klewe}},
  \bibinfo {author} {\bibfnamefont {D.}~\bibnamefont {Meier}}, \bibinfo
  {author} {\bibfnamefont {O.}~\bibnamefont {Manos}}, \bibinfo {author}
  {\bibfnamefont {O.}~\bibnamefont {Kuschel}}, \bibinfo {author} {\bibfnamefont
  {J.}~\bibnamefont {Wollschl{\"a}ger}}, \bibinfo {author} {\bibfnamefont
  {L.}~\bibnamefont {Bouchenoire}}, \bibinfo {author} {\bibfnamefont {S.~D.}\
  \bibnamefont {Brown}}, \bibinfo {author} {\bibfnamefont {J.-M.}\ \bibnamefont
  {Schmalhorst}}, \bibinfo {author} {\bibfnamefont {G.}~\bibnamefont {Reiss}},
  \emph {et~al.},\ }\bibfield  {title} {\enquote {\bibinfo {title}
  {Quantitative disentanglement of the spin seebeck, proximity-induced, and
  ferromagnetic-induced anomalous nernst effect in normal-metal--ferromagnet
  bilayers},}\ }\href@noop {} {\bibfield  {journal} {\bibinfo  {journal}
  {Physical review letters}\ }\textbf {\bibinfo {volume} {119}},\ \bibinfo
  {pages} {227205} (\bibinfo {year} {2017})}\BibitemShut {NoStop}%
\bibitem [{\citenamefont {Altendorf}\ \emph {et~al.}(2011)\citenamefont
  {Altendorf}, \citenamefont {Efimenko}, \citenamefont {Oliana}, \citenamefont
  {Kierspel}, \citenamefont {Rata},\ and\ \citenamefont
  {Tjeng}}]{altendorf2011oxygen}%
  \BibitemOpen
  \bibfield  {author} {\bibinfo {author} {\bibfnamefont {S.}~\bibnamefont
  {Altendorf}}, \bibinfo {author} {\bibfnamefont {A.}~\bibnamefont {Efimenko}},
  \bibinfo {author} {\bibfnamefont {V.}~\bibnamefont {Oliana}}, \bibinfo
  {author} {\bibfnamefont {H.}~\bibnamefont {Kierspel}}, \bibinfo {author}
  {\bibfnamefont {A.}~\bibnamefont {Rata}}, \ and\ \bibinfo {author}
  {\bibfnamefont {L.}~\bibnamefont {Tjeng}},\ }\bibfield  {title} {\enquote
  {\bibinfo {title} {Oxygen off-stoichiometry and phase separation in {E}u{O}
  thin films},}\ }\href@noop {} {\bibfield  {journal} {\bibinfo  {journal}
  {Physical Review B}\ }\textbf {\bibinfo {volume} {84}},\ \bibinfo {pages}
  {155442} (\bibinfo {year} {2011})}\BibitemShut {NoStop}%
\bibitem [{\citenamefont {Steeneken}(2012)}]{steeneken2012new}%
  \BibitemOpen
  \bibfield  {author} {\bibinfo {author} {\bibfnamefont {P.~G.}\ \bibnamefont
  {Steeneken}},\ }\bibfield  {title} {\enquote {\bibinfo {title} {New light on
  {E}u{O} thin films: Preparation, transport, magnetism and spectroscopy of a
  ferromagnetic semiconductor},}\ }\href@noop {} {\bibfield  {journal}
  {\bibinfo  {journal} {arXiv preprint arXiv:1203.6771}\ } (\bibinfo {year}
  {2012})}\BibitemShut {NoStop}%
\bibitem [{\citenamefont {Luo}\ \emph {et~al.}(2008)\citenamefont {Luo},
  \citenamefont {Wang}, \citenamefont {Sun}, \citenamefont {Zhu},\ and\
  \citenamefont {Song}}]{luo2008critical}%
  \BibitemOpen
  \bibfield  {author} {\bibinfo {author} {\bibfnamefont {X.}~\bibnamefont
  {Luo}}, \bibinfo {author} {\bibfnamefont {B.}~\bibnamefont {Wang}}, \bibinfo
  {author} {\bibfnamefont {Y.}~\bibnamefont {Sun}}, \bibinfo {author}
  {\bibfnamefont {X.}~\bibnamefont {Zhu}}, \ and\ \bibinfo {author}
  {\bibfnamefont {W.}~\bibnamefont {Song}},\ }\bibfield  {title} {\enquote
  {\bibinfo {title} {Critical behavior of double perovskite
  {L}a$_2${N}i{M}no$_6$},}\ }\href@noop {} {\bibfield  {journal} {\bibinfo
  {journal} {Journal of Physics: Condensed Matter}\ }\textbf {\bibinfo {volume}
  {20}},\ \bibinfo {pages} {465211} (\bibinfo {year} {2008})}\BibitemShut
  {NoStop}%
\bibitem [{\citenamefont {Stanley}(1971)}]{stanley1971phase}%
  \BibitemOpen
  \bibfield  {author} {\bibinfo {author} {\bibfnamefont {H.~E.}\ \bibnamefont
  {Stanley}},\ }\href@noop {} {\emph {\bibinfo {title} {Phase transitions and
  critical phenomena}}}\ (\bibinfo  {publisher} {Clarendon Press, Oxford},\
  \bibinfo {year} {1971})\BibitemShut {NoStop}%
\bibitem [{\citenamefont {Gel'd}(1975)}]{gel1975critical}%
  \BibitemOpen
  \bibfield  {author} {\bibinfo {author} {\bibfnamefont {P.}~\bibnamefont
  {Gel'd}},\ }\bibfield  {title} {\enquote {\bibinfo {title} {Critical
  phenomena in {E}u{O}},}\ }\href@noop {} {\bibfield  {journal} {\bibinfo
  {journal} {Zh. Eksp. Teor. Fiz}\ }\textbf {\bibinfo {volume} {69}},\ \bibinfo
  {pages} {565--571} (\bibinfo {year} {1975})}\BibitemShut {NoStop}%
\bibitem [{\citenamefont {Idzuchi}\ \emph {et~al.}(2014)\citenamefont
  {Idzuchi}, \citenamefont {Fukuma}, \citenamefont {Park}, \citenamefont
  {Matsuda}, \citenamefont {Tanigaki}, \citenamefont {Aizawa}, \citenamefont
  {Shirai}, \citenamefont {Shindo},\ and\ \citenamefont
  {Otani}}]{idzuchi2014critical}%
  \BibitemOpen
  \bibfield  {author} {\bibinfo {author} {\bibfnamefont {H.}~\bibnamefont
  {Idzuchi}}, \bibinfo {author} {\bibfnamefont {Y.}~\bibnamefont {Fukuma}},
  \bibinfo {author} {\bibfnamefont {H.~S.}\ \bibnamefont {Park}}, \bibinfo
  {author} {\bibfnamefont {T.}~\bibnamefont {Matsuda}}, \bibinfo {author}
  {\bibfnamefont {T.}~\bibnamefont {Tanigaki}}, \bibinfo {author}
  {\bibfnamefont {S.}~\bibnamefont {Aizawa}}, \bibinfo {author} {\bibfnamefont
  {M.}~\bibnamefont {Shirai}}, \bibinfo {author} {\bibfnamefont
  {D.}~\bibnamefont {Shindo}}, \ and\ \bibinfo {author} {\bibfnamefont
  {Y.}~\bibnamefont {Otani}},\ }\bibfield  {title} {\enquote {\bibinfo {title}
  {Critical exponents and domain structures of magnetic semiconductor {E}u{S}
  and gd-doped {E}u{S} films near curie temperature},}\ }\href@noop {}
  {\bibfield  {journal} {\bibinfo  {journal} {Applied Physics Express}\
  }\textbf {\bibinfo {volume} {7}},\ \bibinfo {pages} {113002} (\bibinfo {year}
  {2014})}\BibitemShut {NoStop}%
\bibitem [{\citenamefont {Limmer}\ \emph {et~al.}(2006)\citenamefont {Limmer},
  \citenamefont {Glunk}, \citenamefont {Daeubler}, \citenamefont {Hummel},
  \citenamefont {Schoch}, \citenamefont {Sauer}, \citenamefont {Bihler},
  \citenamefont {Huebl}, \citenamefont {Brandt},\ and\ \citenamefont
  {Goennenwein}}]{limmer2006angle}%
  \BibitemOpen
  \bibfield  {author} {\bibinfo {author} {\bibfnamefont {W.}~\bibnamefont
  {Limmer}}, \bibinfo {author} {\bibfnamefont {M.}~\bibnamefont {Glunk}},
  \bibinfo {author} {\bibfnamefont {J.}~\bibnamefont {Daeubler}}, \bibinfo
  {author} {\bibfnamefont {T.}~\bibnamefont {Hummel}}, \bibinfo {author}
  {\bibfnamefont {W.}~\bibnamefont {Schoch}}, \bibinfo {author} {\bibfnamefont
  {R.}~\bibnamefont {Sauer}}, \bibinfo {author} {\bibfnamefont
  {C.}~\bibnamefont {Bihler}}, \bibinfo {author} {\bibfnamefont
  {H.}~\bibnamefont {Huebl}}, \bibinfo {author} {\bibfnamefont
  {M.}~\bibnamefont {Brandt}}, \ and\ \bibinfo {author} {\bibfnamefont
  {S.}~\bibnamefont {Goennenwein}},\ }\bibfield  {title} {\enquote {\bibinfo
  {title} {Angle-dependent magnetotransport in cubic and tetragonal
  ferromagnets: Application to (001)-and (113) a-oriented (ga, mn) as},}\
  }\href@noop {} {\bibfield  {journal} {\bibinfo  {journal} {Physical Review
  B}\ }\textbf {\bibinfo {volume} {74}},\ \bibinfo {pages} {205205} (\bibinfo
  {year} {2006})}\BibitemShut {NoStop}%
\bibitem [{\citenamefont {Uchida}\ \emph {et~al.}(2015)\citenamefont {Uchida},
  \citenamefont {Qiu}, \citenamefont {Kikkawa}, \citenamefont {Iguchi},\ and\
  \citenamefont {Saitoh}}]{uchida2015spin}%
  \BibitemOpen
  \bibfield  {author} {\bibinfo {author} {\bibfnamefont {K.-i.}\ \bibnamefont
  {Uchida}}, \bibinfo {author} {\bibfnamefont {Z.}~\bibnamefont {Qiu}},
  \bibinfo {author} {\bibfnamefont {T.}~\bibnamefont {Kikkawa}}, \bibinfo
  {author} {\bibfnamefont {R.}~\bibnamefont {Iguchi}}, \ and\ \bibinfo {author}
  {\bibfnamefont {E.}~\bibnamefont {Saitoh}},\ }\bibfield  {title} {\enquote
  {\bibinfo {title} {Spin hall magnetoresistance at high temperatures},}\
  }\href@noop {} {\bibfield  {journal} {\bibinfo  {journal} {Applied Physics
  Letters}\ }\textbf {\bibinfo {volume} {106}},\ \bibinfo {pages} {052405}
  (\bibinfo {year} {2015})}\BibitemShut {NoStop}%
\bibitem [{\citenamefont {Marmion}\ \emph {et~al.}(2014)\citenamefont
  {Marmion}, \citenamefont {Ali}, \citenamefont {McLaren}, \citenamefont
  {Williams},\ and\ \citenamefont {Hickey}}]{marmion2014temperature}%
  \BibitemOpen
  \bibfield  {author} {\bibinfo {author} {\bibfnamefont {S.}~\bibnamefont
  {Marmion}}, \bibinfo {author} {\bibfnamefont {M.}~\bibnamefont {Ali}},
  \bibinfo {author} {\bibfnamefont {M.}~\bibnamefont {McLaren}}, \bibinfo
  {author} {\bibfnamefont {D.}~\bibnamefont {Williams}}, \ and\ \bibinfo
  {author} {\bibfnamefont {B.}~\bibnamefont {Hickey}},\ }\bibfield  {title}
  {\enquote {\bibinfo {title} {Temperature dependence of spin {H}all
  magnetoresistance in thin {YIG/Pt} films},}\ }\href@noop {} {\bibfield
  {journal} {\bibinfo  {journal} {Physical Review B}\ }\textbf {\bibinfo
  {volume} {89}},\ \bibinfo {pages} {220404} (\bibinfo {year}
  {2014})}\BibitemShut {NoStop}%
\bibitem [{\citenamefont {Isasa}\ \emph {et~al.}(2015)\citenamefont {Isasa},
  \citenamefont {Villamor}, \citenamefont {Hueso}, \citenamefont {Gradhand},\
  and\ \citenamefont {Casanova}}]{isasa2015temperature}%
  \BibitemOpen
  \bibfield  {author} {\bibinfo {author} {\bibfnamefont {M.}~\bibnamefont
  {Isasa}}, \bibinfo {author} {\bibfnamefont {E.}~\bibnamefont {Villamor}},
  \bibinfo {author} {\bibfnamefont {L.~E.}\ \bibnamefont {Hueso}}, \bibinfo
  {author} {\bibfnamefont {M.}~\bibnamefont {Gradhand}}, \ and\ \bibinfo
  {author} {\bibfnamefont {F.}~\bibnamefont {Casanova}},\ }\bibfield  {title}
  {\enquote {\bibinfo {title} {Temperature dependence of spin diffusion length
  and spin {H}all angle in {A}u and {P}t},}\ }\href@noop {} {\bibfield
  {journal} {\bibinfo  {journal} {Physical Review B}\ }\textbf {\bibinfo
  {volume} {91}},\ \bibinfo {pages} {024402} (\bibinfo {year}
  {2015})}\BibitemShut {NoStop}%
\bibitem [{\citenamefont {Okamoto}(2016)}]{okamoto2016spin}%
  \BibitemOpen
  \bibfield  {author} {\bibinfo {author} {\bibfnamefont {S.}~\bibnamefont
  {Okamoto}},\ }\bibfield  {title} {\enquote {\bibinfo {title} {Spin injection
  and spin transport in paramagnetic insulators},}\ }\href@noop {} {\bibfield
  {journal} {\bibinfo  {journal} {Physical Review B}\ }\textbf {\bibinfo
  {volume} {93}},\ \bibinfo {pages} {064421} (\bibinfo {year}
  {2016})}\BibitemShut {NoStop}%
\bibitem [{\citenamefont {Cornelissen}\ \emph {et~al.}(2016)\citenamefont
  {Cornelissen}, \citenamefont {Peters}, \citenamefont {Bauer}, \citenamefont
  {Duine},\ and\ \citenamefont {van Wees}}]{cornelissen2016magnon}%
  \BibitemOpen
  \bibfield  {author} {\bibinfo {author} {\bibfnamefont {L.~J.}\ \bibnamefont
  {Cornelissen}}, \bibinfo {author} {\bibfnamefont {K.~J.}\ \bibnamefont
  {Peters}}, \bibinfo {author} {\bibfnamefont {G.~E.}\ \bibnamefont {Bauer}},
  \bibinfo {author} {\bibfnamefont {R.}~\bibnamefont {Duine}}, \ and\ \bibinfo
  {author} {\bibfnamefont {B.~J.}\ \bibnamefont {van Wees}},\ }\bibfield
  {title} {\enquote {\bibinfo {title} {Magnon spin transport driven by the
  magnon chemical potential in a magnetic insulator},}\ }\href@noop {}
  {\bibfield  {journal} {\bibinfo  {journal} {Physical Review B}\ }\textbf
  {\bibinfo {volume} {94}},\ \bibinfo {pages} {014412} (\bibinfo {year}
  {2016})}\BibitemShut {NoStop}%
\bibitem [{\citenamefont {V{\'e}lez}\ \emph {et~al.}(2018)\citenamefont
  {V{\'e}lez}, \citenamefont {Golovach}, \citenamefont {Gomez-Perez},
  \citenamefont {Bui}, \citenamefont {Rivadulla}, \citenamefont {Hueso},
  \citenamefont {Bergeret},\ and\ \citenamefont {Casanova}}]{velez2018spin}%
  \BibitemOpen
  \bibfield  {author} {\bibinfo {author} {\bibfnamefont {S.}~\bibnamefont
  {V{\'e}lez}}, \bibinfo {author} {\bibfnamefont {V.~N.}\ \bibnamefont
  {Golovach}}, \bibinfo {author} {\bibfnamefont {J.~M.}\ \bibnamefont
  {Gomez-Perez}}, \bibinfo {author} {\bibfnamefont {C.~T.}\ \bibnamefont
  {Bui}}, \bibinfo {author} {\bibfnamefont {F.}~\bibnamefont {Rivadulla}},
  \bibinfo {author} {\bibfnamefont {L.~E.}\ \bibnamefont {Hueso}}, \bibinfo
  {author} {\bibfnamefont {F.~S.}\ \bibnamefont {Bergeret}}, \ and\ \bibinfo
  {author} {\bibfnamefont {F.}~\bibnamefont {Casanova}},\ }\bibfield  {title}
  {\enquote {\bibinfo {title} {Spin-hall magnetoresistance in a low-dimensional
  magnetic insulator},}\ }\href@noop {} {\bibfield  {journal} {\bibinfo
  {journal} {arXiv preprint arXiv:1805.11225}\ } (\bibinfo {year}
  {2018})}\BibitemShut {NoStop}%
\end{thebibliography}
%

\end{document}